\PassOptionsToPackage{dvipsnames,table}{xcolor}
\PassOptionsToPackage{unicode=true}{hyperref}
\documentclass[a4paper,longbibliography,english,reprint,aps,superscriptaddress,nofootinbib]{revtex4-2}
\usepackage{makecell}
\usepackage{array}
\usepackage{tabularx}
\usepackage{amssymb}
\usepackage{upgreek}
\usepackage{babel,calc,amsmath,amsthm,bm,graphicx,siunitx}
\usepackage{booktabs}

\graphicspath{{images/}{images/SM/}}

\usepackage[T1]{fontenc}
\usepackage{lmodern}
\makeatletter
\AtBeginDocument{\def\NAT@cmprs{\z@}}
\makeatother
\usepackage{mathdots}
\usepackage{geometry}
\usepackage{hyperref}
\usepackage{mathtools}
\renewcommand{\arraystretch}{1.2}
\usepackage{multirow}
\usepackage{lipsum}
\usepackage[normalem]{ulem}

\usepackage{arydshln}

\usepackage{xcolor}
\usepackage{ragged2e}
\usepackage{tabularray}
\UseTblrLibrary{booktabs}
\usepackage{colortbl}

\definecolor{headerblue}{HTML}{1F4E79}
\definecolor{catblue}{HTML}{D6E4F0}
\definecolor{rowalt}{HTML}{F2F7FB}

\usepackage[ruled]{algorithm2e}
\SetKwInput{KwData}{Input}
\SetKwInput{KwResult}{Output}
\SetKwComment{Comment}{~\#~}{}

\hypersetup{
     colorlinks=true,               
     linkcolor=Blue,            
     citecolor=Blue,             
     urlcolor=Blue,          
 }

\newtheorem*{corollary*}{Corollary}

\newtheorem*{proposition*}{Proposition}

\newtheorem*{theorem*}{Theorem}

\theoremstyle{definition}

\newtheorem*{definition*}{Definition}

\theoremstyle{remark}

\newtheorem*{lemma*}{Lemma}

\newtheorem*{remark*}{Remark}

\newtheorem*{example*}{Example}

\newcommand{\rom}[1]{\MakeUppercase{\romannumeral #1}}

\newif\ifdebug

\debugtrue

\ifdebug

\newcommand\delete{\bgroup\markoverwith{\textcolor{Maroon}{\rule[0.5ex]{2pt}{0.4pt}}}\ULon}

\else

\newcommand{\note}[1]{\ignorespaces}
\newcommand{\delete}[1]{\ignorespaces}

\fi

\begin{document}
\renewcommand{\figurename}{Fig.}

\newcommand{\extfig}{Extended Data Figure}
\newcommand{\methodsname}{Methods}
\newcommand{\smname}{Supplementary Material}

\renewcommand{\sectionautorefname}{Sec.}
\renewcommand{\tableautorefname}{Table}
\renewcommand{\equationautorefname}{Eq.}

\title{Four Generations of Quantum Biomedical Sensors}

\author{Xin Jin}
\thanks{These authors contributed equally to this work.}
\affiliation{Department of Computer Science, University of Pittsburgh, Pittsburgh, PA 15260, USA}

\let\oldfootnote\thefootnote
\renewcommand{\thefootnote}{$\P$}
\let\thefootnote\oldfootnote
\author{Priyam Srivastava}
\thanks{These authors contributed equally to this work.}
\affiliation{Department of Informatics and Networked Systems, University of Pittsburgh, Pittsburgh, PA 15260, USA}
\author{Ronghe Wang}
\affiliation{Department of Physics and Astronomy, University of Pittsburgh, Pittsburgh, PA 15260, USA}
\author{Yuqing Li}
\affiliation{Department of Computer Science, University of Pittsburgh, Pittsburgh, PA 15260, USA}
\author{Jonathan Beaumariage}
\affiliation{Department of Physics and Astronomy, University of Pittsburgh, Pittsburgh, PA 15260, USA}
\author{Tom Purdy}
\affiliation{Department of Physics and Astronomy, University of Pittsburgh, Pittsburgh, PA 15260, USA}
\author{M. V. Gurudev Dutt}
\affiliation{Department of Physics and Astronomy, University of Pittsburgh, Pittsburgh, PA 15260, USA}
\author{Kang Kim}
\affiliation{Department of Medicine, University of Pittsburgh School of Medicine, Pittsburgh, PA 15213, USA}
\affiliation{Department of Bioengineering, University of Pittsburgh, Pittsburgh, PA 15261, USA}
\author{Kaushik Seshadreesan}
\affiliation{Department of Informatics and Networked Systems, University of Pittsburgh, Pittsburgh, PA 15260, USA}
\affiliation{Department of Physics and Astronomy, University of Pittsburgh, Pittsburgh, PA 15260, USA}

\author{Junyu Liu}
\email[Corresponding author: ]{junyuliu@pitt.edu}
\affiliation{Department of Computer Science, University of Pittsburgh, Pittsburgh, PA 15260, USA}

\date{\today}

\begin{abstract}
    Quantum sensing technologies offer transformative potential for ultra-sensitive biomedical sensing, yet their clinical translation remains constrained by classical noise limits and a reliance on macroscopic ensembles. We propose a unifying generational framework to organize the evolving landscape of quantum biosensors based on their utilization of quantum resources. First-generation devices utilize discrete energy levels for signal transduction but follow classical scaling laws. Second-generation sensors exploit quantum coherence, extending precision with the coherence time up to the standard quantum limit, while third-generation architectures employ entanglement and spin squeezing to approach Heisenberg-limited precision. We define an emerging fourth generation characterized by the end-to-end integration of quantum sensing with quantum learning and variational circuits, enabling adaptive inference directly within the quantum domain. By introducing a bandwidth-matching analysis pairing the neural signal hierarchy with platform response bandwidths, classifying deployed clinical devices by precision-scaling class and sensor-tissue proximity, and outlining a staged physical-milestone roadmap toward learning-integrated sensor networks, we identify key technological bottlenecks and chart the transition from measuring physical observables to extracting structured biological information with quantum-enhanced intelligence.
\end{abstract}

\maketitle

\section{Introduction}
Quantum sensing has emerged as a transformative paradigm in measurement science, exploiting the fundamental principles of quantum mechanics to detect physical quantities with unprecedented sensitivity and precision~\cite{RevModPhys.89.035002, giovannetti2004quantum, aslam2023quantum, bakhshandeh2022quantum}. In contrast to conventional biomedical sensing, which relies on macroscopic transduction and ensemble-averaged signals limited by classical noise, quantum sensors encode physical parameters into quantum states whose evolution can be measured with enhanced precision~\cite{giovannetti2011advances}. By harnessing quantum superposition, coherence, and entanglement, quantum sensing enables detection of extremely weak magnetic and electric fields, temperature variations, pressure fluctuations, and chemical or molecular environments~\cite{zhang2021toward}. These capabilities allow sensitivities that surpass the standard quantum limit (SQL) and in principle approach the Heisenberg limit~\cite{zwierz2012ultimate}. Classical biomedical sensors ultimately depend on quantum physics at the microscopic level, but they do not deliberately manipulate quantum resources such as coherence or entanglement. Quantum sensors, in contrast, explicitly engineer quantum states as metrological resources to enhance measurement precision.

\begin{figure}[htb!]
\centering
\includegraphics[width=\columnwidth]{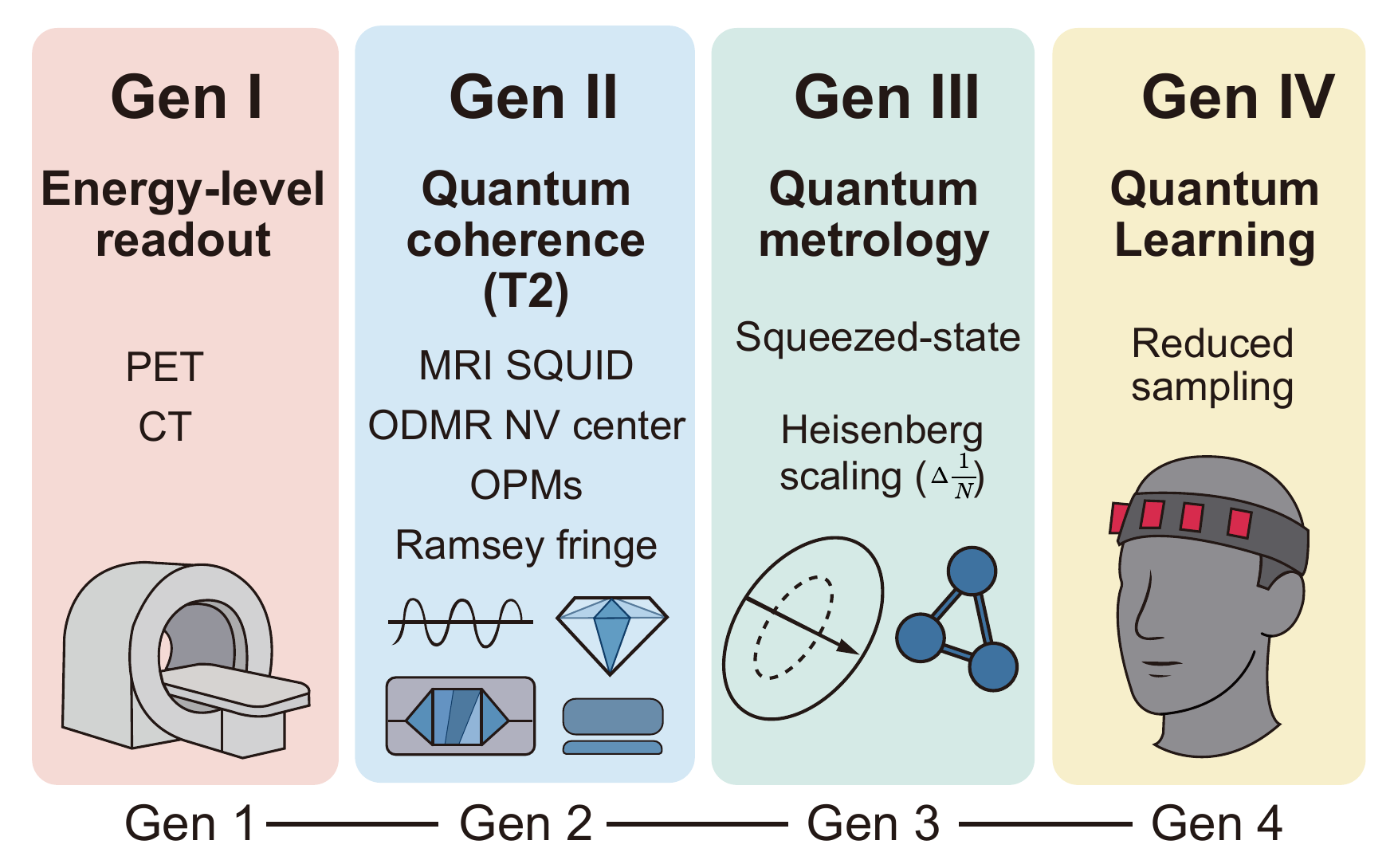}
\caption{Conceptual generational map illustrating the technological evolution of quantum biosensing. The diagram delineates three overarching paradigms (quantum spectroscopy, quantum metrology, and quantum learning), each enabling higher sensitivity and functional integration. The 1st-4th generation biosensors represent successive advances from basic spin readout to coherence-based sensing, entanglement-enhanced metrology, and emerging quantum-AI co-designed medical schemes (Appendix \ref{appendix:timeline}).}
\label{fig:timeline}
\end{figure}

\textbf{The Importance of a Generational Framework.} Rapid advances in biomedical quantum sensing have produced a diverse landscape of sensing platforms, including superconducting circuits, nitrogen-vacancy (NV) centers, atomic magnetometers, and entanglement-enhanced metrology. Despite this progress, the field lacks a systematic framework for comparing sensing architectures according to the quantum resources they employ. Existing overviews organize the field by sensing platform and accessible length scale~\cite{aslam2023quantum}, by definitional resource criteria for quantum sensing at large~\cite{RevModPhys.89.035002}, or by exemplar application domains and the institutional path to clinical translation~\cite{LoFiego2025}; none orders biomedical sensing architectures by the quantum resource entering the measurement chain and the precision scaling it confers. Here we propose a generational framework for biomedical quantum sensors that organizes sensing technologies according to the depth of quantum resource utilization. This framework provides conceptual clarity, establishes consistent benchmarks for technological maturity, and identifies the natural progression from physical signal detection toward intelligent quantum-enhanced inference.

\textbf{Current Landscape of Quantum Biomedical Sensors.} Modern biomedical measurement stands at a crossroads between established clinical modalities and emerging technologies that exploit genuine quantum phenomena like coherence and entanglement. Superconducting quantum interference devices (SQUIDs) remain the most mature platform, providing femtotesla ($10^{-15}$ T) sensitivity essential for detecting neural and cardiac fields in magnetoencephalography (MEG) and magnetocardiography (MCG)~\cite{fenici2025advanced}. Beyond SQUIDs, other coherence-based sensors, such as solid-state NV centers and cold-atom interferometers, have expanded the scope of biometrology to the nanoscale, enabling high-sensitivity, real-time functional readouts~\cite{barry2016optical, taylor2008high}. Entanglement-enhanced sensing has further demonstrated precision improvements beyond classical limits in solid-state spin platforms~\cite{Zhou2025,Rovny2025}. These developments signal a broader shift: moving from simple physical observations toward hybrid, learning-enabled systems that integrate quantum resources with computational intelligence (Fig.~\ref{fig:timeline}).

We delineate four conceptual generations of biomedical quantum sensors, categorized by their utilization of quantum resources and the depth of information integration:

\begin{itemize}
\item[(\rom{1})] \textbf{Gen~\rom{1}: Energy-Level Readout.} 
These sensors utilize discrete quantum energy levels as sensing elements but rely on population or spectral readout without maintaining coherent superposition. Operating effectively as classical sensors with quantum-scale transduction, this generation includes magnetoresistive sensors, quantum dots, and fluorescence spectroscopy~\cite{RevModPhys.89.035002}.

\item[(\rom{2})] \textbf{Gen~\rom{2}: Quantum Coherence.} 
This stage exploits quantum coherence such as wave-like spatial or temporal superposition states to measure physical quantities. Achievable precision scales with the coherence time ($T_2$), a principle demonstrated in MRI as well as solid-state systems like NV centers and atomic interferometers.

\item[(\rom{3})] \textbf{Gen~\rom{3}: Quantum Metrology.} 
Third-generation architectures leverage quantum entanglement and spin squeezing to improve sensitivity beyond classical limits. By employing entangled or correlated ensembles, these systems enable information encoding that approaches the Heisenberg limit.

\item[(\rom{4})] \textbf{Gen~\rom{4}: Quantum Learning.}
The emerging frontier integrates quantum sensors with quantum or hybrid quantum-classical processors to process measured information adaptively. This paradigm enables data-driven, intelligent sensing and real-time interpretation of biological signals, marking a transition from raw measurement to active learning.
\end{itemize}

This framework formalizes the evolution of biomedical sensing from observing physical observables to interpreting structured biological information through quantum-enhanced intelligence.
The remainder of this Perspective develops the proposed generational framework for biomedical quantum sensing. Section~II reviews the foundations of biomedical measurement and the motivations for quantum sensing. Section~III introduces the four sensor generations. Section~IV discusses key physical constraints across platforms. The final section outlines future directions toward clinical translation.
\begin{figure*}[htb!]
\centering
\includegraphics[width=\textwidth]{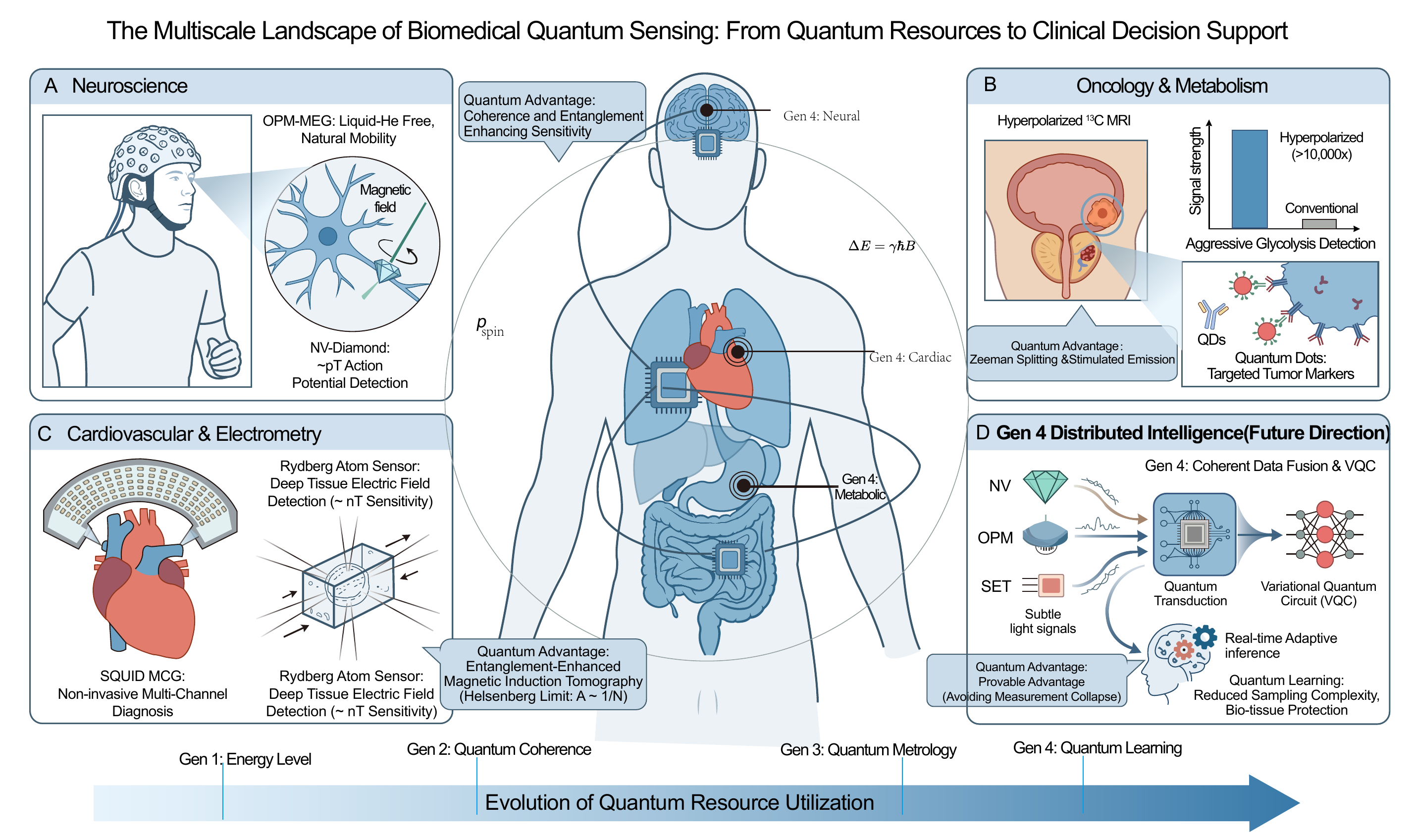}
\caption{Multiscale applications and generational evolution of biomedical quantum sensing. \textbf{A-C,} Representative clinical deployments across neuroscience, oncology, and cardiovascular medicine, highlighting quantum advantages in Gen 1-3 platforms. \textbf{D,} Emerging paradigm of Gen 4 distributed intelligence, featuring coherent data fusion via quantum transduction and adaptive inference through variational quantum circuits (VQC). The bottom axis illustrates the transition from discrete energy-level readout (Gen 1) to advanced quantum learning (Gen 4).}

\end{figure*}

\section{Related Work}

Medical imaging technologies have transformed modern medicine by 
enabling non-invasive visualization of anatomical, functional, 
and metabolic processes, beginning with 
X-rays in 1895~\cite{rontgen1895xray}, followed by 
computed tomography (CT)~\cite{hounsfield1973computerized} and 
positron emission tomography (PET)~\cite{ter1975positron}. CT and PET rely strictly on classical radiation detection and 
are not classified as quantum sensors. Among established clinical 
technologies, magnetic resonance imaging 
(MRI)~\cite{lauterbur1973image,mansfield1977multi} operates on
coherent precession of nuclear spins, designating it as a second-generation modality within the 
generational framework developed here.

Despite their tremendous clinical impact, these imaging modalities 
rely on fundamentally classical measurement processes whose precision 
is ultimately constrained by stochastic noise sources. Specifically, 
thermal noise described by the Johnson-Nyquist 
theorem~\cite{johnson1928thermal,nyquist1928thermal} sets the minimum 
detectable signal in resistive systems, while photon shot noise 
fundamentally limits optical and nuclear 
imaging~\cite{mandel1995optical}. Together with measurement limits 
such as the SQL~\cite{caves1981quantum}, these noise sources impose rigid 
barriers to sensitivity enhancement. These limitations are 
particularly problematic in applications requiring the detection of 
extremely weak or highly localized biological signals, including 
early-stage disease biomarkers~\cite{hanash2008mining}, single-cell 
processes~\cite{altschuler2010cellular}, and rapid physiological 
dynamics~\cite{orphanidou2019review}. Representative examples include 
neural magnetic fields at the femtotesla level at the scalp, weak molecular signals
in early disease stages, and transient physiological processes 
occurring on millisecond timescales. Beyond sensitivity limitations, 
classical medical imaging modalities also face important practical 
constraints, including ionizing radiation exposure in X-ray and 
CT~\cite{brenner2007computed}, toxicity risks of gadolinium-based 
contrast agents required for oncological and vascular 
MRI~\cite{caravan1999gadolinium}, limited penetration depth in 
optical imaging~\cite{ntziachristos2010going}, and restricted temporal 
resolution in metabolic imaging~\cite{sorenson1987physics}. These 
challenges have motivated quantum sensing as a promising 
approach to access elusive physiological signals. Unlike 
classical energy-based measurements, quantum sensors encode physical 
parameters into coherent phase evolution of quantum states via 
quantum superposition~\cite{giovannetti2006quantum}, while 
quantum entanglement enables correlated 
measurements surpassing classical precision 
bounds~\cite{giovannetti2011advances}. With the quantum Cram\'{e}r-Rao 
bound~\cite{helstrom1967minimum} defining ultimate precision
limits, these principles establish the
foundation for quantum-enhanced biomedical sensing.

Biomedical environments present one of the most demanding operating regimes 
for quantum sensors, fundamentally mismatched with the isolation and stability 
typically required for quantum-enhanced measurements. Living biological systems 
operate in warm, hydrated, and dynamically fluctuating environments that 
induce rapid decoherence of fragile quantum states~\cite{schlosshauer2007decoherence,
taylor2016quantum}. As a result,
achievable sensitivity, coherence time, and sensing distance are jointly 
constrained by the physical properties of biological tissue rather than by 
quantum resources alone~\cite{RevModPhys.89.035002}. Several dominant noise 
sources further limit practical performance in clinical settings. 
Electromagnetic interference from medical instrumentation, power systems, and 
surrounding electronics introduces broadband background noise, while 
mechanical motion arising from respiration, cardiac activity, and involuntary 
patient movement produces time-varying artifacts that degrade signal stability. 
In addition, magnetic fluctuations from nearby ferromagnetic materials and 
structural components can distort weak biomagnetic 
signals~\cite{boto2018moving}, imposing stringent requirements on shielding 
and sensor placement~\cite{tierney2019optically}. Beyond environmental noise, 
translational and clinical constraints impose additional limitations. Many 
high-performance quantum sensors, such as SQUIDs, rely on cryogenic temperatures and bulky dewar 
systems~\cite{hamalainen1993magnetoencephalography}. While newer modalities 
such as optically pumped magnetometers and NV centers operate at or near room 
temperature, they still face challenges related to intense optical fields or 
microwave radiation that may lead to localized tissue heating~\cite{wu2016diamond}. 
Furthermore, biomedical deployment requires strict biocompatibility of sensor 
packaging~\cite{ratner2004biomaterials,schirhagl2014nitrogen}, non-invasive 
operation, and adherence to safety guidelines regarding electromagnetic
exposure~\cite{international2020guidelines} and, for optically interrogated
platforms, laser-exposure limits~\cite{icnirp2013laser}. Consequently, realizing the full potential of biomedical quantum sensing demands the successful integration of intrinsic quantum sensitivity with clinically practical, biologically compatible system designs and workflows.

\textbf{Relation to existing frameworks.} Several organizing schemes for this field already exist. Degen, Reinhard, and Cappellaro classify quantum sensors by definitional criteria (Types I-III)~\cite{RevModPhys.89.035002}: a foundational trichotomy that is static rather than developmental, is not mapped onto biomedical deployment, and contains no learning tier. Aslam et al.\ survey biomedical quantum sensing by platform and accessible length scale~\cite{aslam2023quantum}; their outlook centers on sensitivity gains through materials and protocols rather than a pathway from standard-quantum-limit operation to entanglement-enhanced sensing in biological settings. Most recently, Lo Fiego et al.\ highlight the near-term clinical promise of quantum biosensing through three exemplar domains and argue that the decisive barriers to translation are structural and societal as much as technical~\cite{LoFiego2025}; the generational succession they describe, from quantum dots to spin-modulated nanodiamonds to emerging molecular spin systems, is a succession of label materials within a single application domain. The framework developed here takes a different, complementary axis: generations are keyed to the quantum resource entering the measurement chain and the precision scaling it confers, extended to a learning tier; where these works map platforms, materials, and the institutional path to the clinic, we map the physical-resource path and its bottleneck-by-bottleneck clinical payoff.

\section{Quantum Medical Sensor Generations}

\definecolor{TblHead}{HTML}{F5ECD7}   
\definecolor{SecRow}{HTML}{FAE5C8}    
\definecolor{Stripe}{HTML}{FEFEF5}    

\begin{table*}[htbp]
\centering\small
\caption{Unified comparison of representative first-, second-, and third-generation quantum sensing platforms for biomedical applications.
Resolution denotes intrinsic detection capability for 1st/2nd-generation sensors; Gain quantifies the demonstrated improvement beyond the standard quantum limit~(SQL) for 3rd-generation platforms.
Entanglement resources employed by third-generation sensors are indicated by superscript footnotes ($^{a}$-$^{i}$).
Biomedical viability is jointly constrained by operating temperature~$T_{\mathrm{op}}$, bio-sample thermal compatibility~$T_{\mathrm{bio}}$, and sensor-sample distance. $T_{\mathrm{op}}$ denotes the temperature that must be actively maintained for the sensing element or its enabling hardware (parenthetical notes identify the cryogenic component), a proxy for the infrastructure burden that limits mobility; Distance denotes the sensor-sample operating standoff during measurement, not the imaging depth accessible within the sample. $T_{\mathrm{bio}}$ reflects the thermal environment experienced by the biological sample, determined jointly by $T_{\mathrm{op}}$ and the sensor-sample distance: when the sensor operates in the near field with strong thermal coupling, $T_{\mathrm{bio}} \approx T_{\mathrm{op}}$; when sufficient standoff distance provides thermal isolation, $T_{\mathrm{bio}}$ remains at physiological temperature (${\sim}\,300$\,K).}
\label{tab:merged_quantum_sensors}

\resizebox{\textwidth}{!}{%
\begin{tblr}{
  colspec  = {l c l l l l l},
  rowsep   = 1.5pt,
  colsep   = 5pt,
  hline{1,Z} = {0.08em},
  hline{2}   = {0.05em},
  row{1}  = {bg=TblHead, font=\bfseries\small},
  row{2}  = {bg=SecRow, font=\itshape\small},
  row{10} = {bg=SecRow, font=\itshape\small},
  row{20} = {bg=SecRow, font=\itshape\small},
  row{4,6,8,12,14,16,18,22,24,26,28} = {bg=Stripe},
}
\textbf{Platform} &
\textbf{Gen.} &
\textbf{Measured quantity} &
\textbf{Resolution / Gain} &
$\bm{T_{\mathrm{op}}}$ &
$\bm{T_{\mathrm{bio}}}$ &
\textbf{Distance} \\
\SetCell[c=7]{l} \textit{A.\enspace First Generation: Energy-Level Readout} & & & & & & \\
GMR sensors                & 1st & Magnetoresistance           & nT magnetic fields             & 233-423\,K   & 298-310\,K & Near-field        \\
TMR sensors                & 1st & Tunneling current           & nT magnetic fields             & 233-398\,K   & 298-310\,K & ${<}\,1$\,mm      \\
Quantum dots / nanowires   & 1st & Optical / electrical signal & Molecular sensitivity          & ${\sim}\,300$\,K  & 293-318\,K & Near-field        \\
SET                        & 1st & Single-electron charge      & Single-charge sensitivity      & 0.01-40\,K   & 0.01-40\,K & ${<}\,100$\,nm    \\
NV centers (ensemble)      & 1st & Local magnetic field        & pT-nT magnetic fields         & 288-333\,K   & 293-310\,K & ${<}\,1$\,mm      \\
Muon probes ($\mu$SR)      & 1st & Local magnetic field        & High local sensitivity         & 0.03-1000\,K & 2-300\,K   & Near-field        \\
Neutron probes             & 1st & Magnetic / nuclear contrast & $\upmu$T-nT contrast            & 293-303\,K   & 2-300\,K   & 10\,cm-1\,m      \\
\hline
\SetCell[c=7]{l} \textit{B.\enspace Second Generation: Quantum Coherence} & & & & & & \\
NMR sensors                & 2nd & Magnetic field              & $\propto 1/\!\sqrt{N}$                                       & ${\sim}\,4$\,K (magnet)   & ${\sim}\,300$\,K & 5\,mm-2\,cm      \\
MRI / fMRI                 & 2nd & Spin relaxation             & $0.47\;\mathrm{fT}/\!\sqrt{\mathrm{Hz}}$ (ULF)~\cite{Savukov2022}                     & ${\sim}\,4$\,K (magnet)   & ${\sim}\,300$\,K & 5\,mm-2\,cm      \\
NV centers (Ramsey/echo)   & 2nd & Local B-field (ionic)       & ${<}\,10\;\mathrm{pT}/\!\sqrt{\mathrm{Hz}}$                  & 295-300\,K   & ${\sim}\,300$\,K & ${<}\,100$\,nm-1\,$\upmu$m \\
SQUID                      & 2nd & Magnetic flux               & $\mathrm{fT}$-$\mathrm{pT}/\!\sqrt{\mathrm{Hz}}$           & 4-20\,K      & ${\sim}\,300$\,K & 1-25\,mm         \\
OPMs (SERF / Larmor)       & 2nd & B-field (atomic spin)       & $\mathrm{fT}$-$\mathrm{pT}/\!\sqrt{\mathrm{Hz}}$           & 393-473\,K   & ${\sim}\,300$\,K & 1-5\,mm          \\
Rydberg atom sensors       & 2nd & E-field amplitude           & ${\sim}\,30\;\upmu\mathrm{V\,cm^{-1}}/\!\sqrt{\mathrm{Hz}}$~\cite{sedlacek2012microwave}                  & 298-353\,K   & ${\sim}\,300$\,K & 0.1-1\,mm (thin cell); cm (std.)~\cite{Kbler2010,Zhang2018}  \\
Flux qubits~\cite{toida2023magnetometry}                & 2nd & Magnetic flux var.          & $\mathrm{pT}/\!\sqrt{\mathrm{Hz}}$                           & ${\leq}\,20$\,mK  & ${\leq}\,20$\,mK          & ${\sim}\,1$-$10\;\upmu$m    \\
Charge qubits~\cite{tanarom2025characteristic}              & 2nd & E-field / charge            & $10^{-7}\;e/\!\sqrt{\mathrm{Hz}}$                            & ${\leq}\,100$\,mK & ${\leq}\,100$\,mK         & ${<}\,100$\,nm        \\
Atom interferometers       & 2nd & Inertial phase              & $10^{-9}\;g$                                                 & $\upmu$K-100\,nK      & ${\sim}\,300$\,K & ${<}\,100\,\upmu$m    \\
\hline
\SetCell[c=7]{l} \textit{C.\enspace Third Generation: Entanglement-Enhanced Metrology} & & & & & & \\
NV centers (entangled)$^{a}$~\cite{Zhou2025}                            & 3rd & Magnetic field        & ${\sim}\,5$\,dB   & ${\sim}\,300$\,K  & ${\sim}\,300$\,K  & ${<}\,100$\,nm  \\
NV ensemble (spin-squeezed)$^{b}$~\cite{Wu2025}                         & 3rd & Magnetic field        & ${\sim}\,0.5$\,dB & ${\sim}\,300$\,K  & ${\sim}\,300$\,K  & ${<}\,1$\,mm    \\
Optomechanical systems$^{c}$~\cite{Li:18}                               & 3rd & Force / displacement  & ${\sim}\,2$\,dB   & ${\sim}\,300$\,K  & ${\sim}\,300$\,K  & Near-field  \\
Trapped ions (entangled)$^{d}$~\cite{doi:10.1126/science.1097576}       & 3rd & Frequency / B-field   & 4-9\,dB          & mK regime       & ${\sim}\,300$\,K  & ${\sim}$\,mm    \\
Atomic vapor (spin-squeezed)$^{e}$~\cite{PhysRevLett.130.203602}        & 3rd & Magnetic field        & ${\sim}\,2$\,dB   & 350-470\,K     & ${\sim}\,300$\,K  & mm-cm      \\
Atomic vapor (entangled)$^{f}$~\cite{doi:10.1126/sciadv.adg1760}        & 3rd & Magnetic gradient     & 5.5\,dB           & 350-470\,K     & ${\sim}\,300$\,K  & cm          \\
Rydberg atom arrays$^{g}$~\cite{Bornet2023}                             & 3rd & Metrological phase    & 3.5-4\,dB        & 10-100\,$\upmu$K   & 10-100\,$\upmu$K   & Near-field  \\
Optical interferometers$^{h}$~\cite{li2025phase}                        & 3rd & Optical path phase    & ${\sim}\,3$\,dB   & ${\sim}\,300$\,K  & ${\sim}\,300$\,K  & mm          \\
Superconducting circuits$^{i}$~\cite{PhysRevX.7.041011}                 & 3rd & Magnetic flux / force & ${\sim}\,1$\,dB   & 10-20\,mK      & ${\sim}\,300$\,K  & mm-cm (in-cryostat)     \\
\end{tblr}}

\vspace{4pt}
\raggedright\small
$^{a}$~Bell-state/GHZ (2-3 spins), ${\sim}\,\upmu$s-ms.\quad
$^{b}$~Dipolar spin squeezing (${\sim}\,10^3$ spins), ${\sim}\,\upmu$s.\quad
$^{c}$~Squeezed light + mechanical entanglement, steady-state.\quad
$^{d}$~GHZ/spin squeezing (2-14+ ions), 1\,ms-0.1\,s.\quad
$^{e}$~Collective spin squeezing (${\sim}\,10^{12}$-$10^{13}$ atoms), 1-100\,ms.\quad
$^{f}$~Entangled photon-spin correlations, ${\sim}\,$1-10\,ms.\quad
$^{g}$~Dipolar spin squeezing (${\sim}\,$70-100 atoms), 10-100\,$\upmu$s.\quad
$^{h}$~Photon-pair entanglement (SPDC), ps correlation time.\quad
$^{i}$~Squeezed microwave vacuum, ${\sim}\,\upmu$s.

\end{table*}

\noindent\textbf{First Generation Quantum Medical Sensor: Energy-Level Readout.}\\
First-generation quantum sensors rely on quantized energy spectra and spin-dependent interactions of electronic or nuclear states to map biological signals onto measurable energy shifts.
In contrast to higher-generation sensors, coherence and superposition are not defining components of the sensing process~\cite{RevModPhys.89.035002, aslam2023quantum}.

Despite relying on relatively simple quantum resources, first-generation quantum sensors have achieved widespread clinical deployment and constitute the backbone of modern medical diagnostics. Representative implementations include magnetoresistive (GMR~\cite{baselt1998biosensor} and TMR~\cite{ghemes2023tunnel}), quantum dots~\cite{michalet2005quantum}, and NV centers with ODMR readout utilizing the Zeeman effect~\cite{schirhagl2014nitrogen, rondin2014magnetometry}. Quantum states serve primarily as energy structure enabling signal transduction rather than engineered quantum information carriers, so measurement precision follows classical scaling laws~\cite{giovannetti2004quantum} despite the sensing mechanism being of quantum origin~\cite{RevModPhys.89.035002}.

Magnetoresistive biology quantum sensors~(Appendix~\ref{appendix:magnetoresistive}) mainly consist of giant and tunnel magnetoresistance
devices~(GMR/TMR), which represent early examples of first-generation quantum medical
sensing in which spin-dependent electronic transport enables robust
signal transduction under ambient conditions. In these platforms,
biological events are indirectly encoded through magnetic labels or
local fields that modulate electronic resistance.

A complementary class of first-generation quantum sensors operate at
the nanoscale through quantization of electronic energy levels and
charge. Quantum dots and single-electron~(Appendix~\ref{appendix: quantum dots/SET}) devices leverage discrete
electronic states, Coulomb blockade, or size-dependent optical emission
to detect molecular binding events and biochemical interactions with
high sensitivity. These platforms have proven particularly effective in
fluorescence-based imaging and label-assisted biosensing, where the
stability of quantized states improves signal robustness. However, the
quantum states involved function solely as static energy spectra rather
than dynamically controlled carriers of information, and measurement
precision remains governed by classical statistics.

Nitrogen-vacancy centers in diamond have emerged as a powerful first-generation quantum sensing platform for magnetic imaging at cellular and subcellular length scales. Although NV-based sensors rely on quantum spin properties of point defects in diamond, first-generation implementations operate through ensemble averaging and optically detected magnetic resonance without exploiting inter-qubit entanglement or actively preserved coherence. Wide-field quantum diamond microscopes employing dense NV ensembles have achieved submicron spatial resolution with sensitivities on the order of two hundred nanotesla, enabling magnetic imaging of individual cells labeled with superparamagnetic nanoparticles \cite{glenn2015single}. Earlier demonstrations resolved intracellular magnetite chains in living magnetotactic bacteria with approximately four hundred nanometer spatial resolution, establishing NV centers as a viable modality for noninvasive magnetic imaging in biological systems \cite{le2013optical}. More recent advances have enabled quantitative immunomagnetic imaging of human tumor sections at micrometer-scale resolution, while remaining compatible with conventional histological and fluorescence workflows \cite{chen2022immunomagnetic}. The chemical inertness and biocompatibility of diamond, together with stable room-temperature operation, position NV-based platforms as promising candidates for ex vivo diagnostics and future minimally invasive monitoring applications.

Beyond room-temperature platforms, muon spin spectroscopy ($\mu$SR) and neutron-based probes offer exceptional sensitivity to local magnetic environments and atomic-scale structure, respectively~\cite{hillier2022muon,rauch2015neutron}. However, both techniques require large-scale facilities, including particle accelerators for $\mu$SR and nuclear reactors or spallation sources for neutron scattering, that fundamentally preclude point-of-care or in vivo deployment~\cite{hillier2022muon}. These facility-dependent platforms remain confined to ex situ characterization of biological materials, positioning them as complementary research tools rather than candidates for widespread biomedical quantum sensing.

Although the platforms above span diverse implementations, they share a common architectural limitation: quantum states serve only as passive transduction elements, and measurement precision obeys classical scaling ($\Delta\theta \propto 1/\sqrt{N}$)~\cite{giovannetti2004quantum}, with no gain from coherent interrogation time; in practice, technical and thermal noise keeps these devices above the projection-noise floor. The performance ceiling is set by classical noise statistics rather than quantum bounds. NV centers make this tension especially visible: the same defect system that supports coherent spin manipulation and entanglement-assisted protocols~\cite{schirhagl2014nitrogen} is, in first-generation implementations, reduced to incoherent ensemble readout. Overcoming this classical precision barrier motivates the transition to second-generation sensors, where quantum coherence is deliberately preserved as a metrological resource.

\noindent\textbf{Second Generation Quantum Medical Sensor: Quantum Coherence.}\\
Second-generation quantum sensors utilize the phase evolution of coherent superpositions to approach the SQL, providing a signal-to-noise ratio~(SNR) that scales with $\sqrt{N}$~\cite{giovannetti2004quantum, PhysRevD.23.1693}. The SQL marks the same $1/\sqrt{N}$ statistical scaling that bounds first-generation devices; what distinguishes second-generation sensors is that coherent phase accumulation allows them to saturate this bound, with precision extending with the coherence time $T_2$. Table~\ref{tab:merged_quantum_sensors} highlights that only a small subset of quantum sensors remain biologically viable once thermal and spatial constraints are imposed. Consistent with prior reviews, NMR-based modalities are grouped under second-generation sensors here, though this adopts a broad definition based on quantum coherence; a stricter criterion would distinguish conventional ensemble MRI from coherence-engineered NMR protocols.

Nuclear magnetic resonance (NMR) sensors represent the earliest and most clinically mature class of quantum medical sensors, exploiting coherent precession of thermal nuclear spin ensembles to encode magnetic, chemical, and spatial information. Frequency and phase encoding established MRI as a non-contact modality for anatomical visualization \cite{lauterbur1973image}, while blood-oxygen-level-dependent (BOLD) contrast enabled functional MRI (fMRI), providing an indirect measure of neural activity through susceptibility-induced modulation of transverse ($T_2^*$) relaxation \cite{ogawa1992intrinsic}. NMR offers deep tissue penetration, physiological-temperature operation, and exceptional chemical specificity, supporting clinical adoption from structural imaging to metabolomics and functional brain mapping \cite{Buergel2022}. However, spin coherence is rapidly averaged over macroscopic ensembles and is not preserved as an actively controllable metrological resource, constraining sensitivity and temporal resolution to scales dictated by thermal polarization and relaxation rather than quantum coherence. Consequently, MRI performance improvements have relied predominantly on engineering advances in hardware, pulse sequences, and signal processing rather than coherence-enabled scaling. Extensions such as hyperpolarized $^{13}$C MRI partially overcome polarization limits for real-time metabolic imaging \cite{sushentsev2022hyperpolarised}, highlighting both the power and the remaining boundaries of NMR-based quantum sensing.

Second-generation NV diamond magnetometers extend first-generation sensing by preserving spin coherence during measurement, improving sensitivity from the hundred-nanotesla regime to single-digit $\mathrm{pT}/\!\sqrt{\mathrm{Hz}}$ and enabling label-free detection of intrinsic bioelectric currents. Using a thirteen-micrometer-thick, high-density NV layer with ${\sim}\,15\;\mathrm{pT}/\!\sqrt{\mathrm{Hz}}$ sensitivity, Barry et al.\ directly measured nanotesla-scale magnetic pulses from single action potentials in invertebrate axons with ${\sim}\,30\;\upmu\mathrm{s}$ temporal resolution~\cite{barry2016optical}, while biophysical modeling predicts three-dimensional localization of axon-hillock currents with approximately $70\%$ accuracy~\cite{parashar2020axon}. Recent ensemble advances have further expanded capability: NV sensors with sensing volumes of ${\sim}\,300 \times 100 \times 20\;\upmu\mathrm{m}^{3}$ and ${\sim}\,50\;\mathrm{pT}/\!\sqrt{\mathrm{Hz}}$ sensitivity over $10\;\mathrm{kHz}$ bandwidth, operated with continuous-wave multi-frequency ODMR, have recorded compound action potentials in live mouse brain slices, with tetrodotoxin perfusion abolishing the signal and confirming its neuronal origin~\cite{hansen2023microscopic}. Material-engineering advances, such as $[111]$-oriented bulk diamond achieving $9.4 \pm 0.1\;\mathrm{pT}/\!\sqrt{\mathrm{Hz}}$ in continuous-wave mode~\cite{sekiguchi2024diamond}, further improve the baseline sensitivity, although the generational distinction rests on active coherence manipulation rather than on CW readout alone. These results position NV-diamond magnetometry at the threshold of contact-free, cellular-scale neurophysiology in surface-accessible preparations: unlike penetrating electrodes, NV sensors record without entering the tissue, and unlike fMRI, which infers neural activity indirectly through hemodynamic responses on second timescales, NV magnetometers directly capture electromagnetic signals with microsecond resolution. Beyond magnetometry, the coherence-based paradigm extends to nanoscale nuclear magnetic resonance: single NV centers combined with dynamic decoupling sequences (e.g., XY8-k) that actively preserve spin coherence have enabled NMR spectroscopy of zeptoliter sample volumes~\cite{lovchinsky2016nuclear}, opening routes to label-free monitoring of drug-target binding kinetics and structural characterization of biomolecular complexes under native conditions. This application exemplifies the defining characteristic of second-generation sensing: measurement sensitivity is directly determined by the actively maintained coherence time rather than by ensemble averaging alone. Despite remaining challenges in sensitivity, array scalability, and clinical integration, these coherence-enabled sensors open a path toward applications such as intraoperative functional mapping and high-throughput screening of patient-derived neural organoids.

SQUIDs are among the most clinically mature second-generation quantum sensors, providing unmatched femtotesla-level magnetic field sensitivity through coherence-based phase detection. Their sensitivity derives from the macroscopic quantum coherence of the superconducting condensate: Josephson interference and flux quantization constitute an intrinsically phase-based readout, warranting a second-generation classification even though the coherence is an equilibrium property of the condensate rather than a pulsed resource. Although conventional SQUID systems require cryogenic cooling and impose large sensor-source standoff distances, recent on-scalp implementations place sensors in close proximity to biological tissue, substantially improving spatial resolution and signal-to-noise ratio. On-scalp SQUID magnetoencephalography has revealed early somatosensory responses with $\sim$16~ms latency that were undetectable using conventional MEG, directly demonstrating the diagnostic value of reduced sensor-to-cortex distance \cite{Andersen2020}. 
In cardiology, multi-channel SQUID magnetocardiography has detected nonischemic cardiomyopathy with a positive predictive value of approximately 93\% (sensitivity 0.59, specificity 0.95) in a 209-patient cohort, supporting its role as a rapid, contactless screening complement to echocardiography \cite{brala2023application}. Beyond neural and cardiac applications, SQUID-based magnetoneurography enables noninvasive imaging of spinal and peripheral nerve conduction \cite{adachi2024squid}, while ultra-low-field MRI using SQUID detectors demonstrates the feasibility of hybrid MEG/MRI systems operating at microtesla fields \cite{guo2020squid}. Despite these successes, widespread clinical adoption remains constrained by cryogenic infrastructure, magnetic shielding requirements, and standoff-distance-limited spatial resolution. These constraints position SQUIDs as a benchmark platform that motivates the development of room-temperature second-generation alternatives \cite{fenici2025advanced}.

Second-generation optically pumped magnetometers (OPMs) exploit coherent precession of spin-polarized alkali vapor ensembles to achieve single-digit femtotesla per root hertz sensitivity at room temperature, eliminating the cryogenic standoff of SQUID-based MEG. On-scalp OPM arrays yield three- to fivefold larger neuronal signal amplitudes for superficial cortical sources, a gain that diminishes to roughly twofold for deeper regions, while permitting natural head movement~\cite{brookes2022magnetoencephalography,boto2018moving}. Wearable systems exceeding fifty channels have been deployed in clinical research studies involving infants and epilepsy patients \cite{brookes2022magnetoencephalography}. An eighty-channel dual-axis OPM helmet has further reproduced visual-evoked and gamma-band responses in twenty healthy adults, including participants whose dental metal precluded conventional MEG recordings, demonstrating robustness under realistic clinical conditions \cite{safar2024using}. Beyond neuroimaging, OPM gradiometer arrays enable quantitative magnetorelaxometry of tumor-targeted magnetic nanoparticles: a twenty-two-sensor platform mapped six micrograms of iron across a $12 \times 8~\mathrm{cm}^2$ field of view \cite{jaufenthaler2020quantitative}, while a pulsed-pump OPM design operating in unshielded environments detected 1.4 micrograms of iron in $100~\upmu\mathrm{L}$ volumes with bandwidths exceeding 100~kHz \cite{jaufenthaler2021pulsed}. Despite these advances, clinical translation remains limited by manufacturing scalability, thermal management, and magnetic interference rejection in hospital environments. Nevertheless, room-temperature operation, flexible positioning, and scalable architectures position OPMs as strong candidates for next-generation clinical neuroimaging, a trajectory now well documented~\cite{brookes2022magnetoencephalography,LoFiego2025}.

Rydberg-atom quantum sensors rely critically on vacuum or sealed vapor cells to preserve atomic coherence and enable high-fidelity electrometry. Atoms excited to high principal quantum numbers exhibit extremely large electric dipole moments and polarizabilities, rendering them exquisitely sensitive to external electromagnetic fields but simultaneously highly susceptible to collisional decoherence from background gas molecules. Without such isolation, Rydberg-state lifetimes collapse to nanosecond scales under ambient conditions, eliminating usable coherence and sensing contrast. While these cell-based architectures enable room-temperature operation without cryogenic infrastructure, the requirement for sealed vacuum environments, optical access, and precise field control fundamentally constrains compatibility with in vivo biological settings, where close sensor-tissue proximity, mechanical compliance, and environmental robustness are essential. As a result, although Rydberg sensors represent a powerful platform for quantum electrometry, their reliance on vacuum cells remains a primary obstacle to direct biomedical translation \cite{fan2015atom}.

Similar to Rydberg sensors, solid-state qubits and matter-wave interferometers offer extraordinary precision but face severe translational barriers. Flux qubits are emerging as powerful quantum sensors for biomedical applications, offering sensitivity at the $\mathrm{pT}/\!\sqrt{\mathrm{Hz}}$ level by transducing small flux changes into shifts of the qubit's microwave transition~\cite{toida2023magnetometry}. Meanwhile, superconducting charge qubits function as extreme electrometers. A heavy-shunted fluxonium qubit operated at $1.8\;\mathrm{MHz}$ recently achieved a charge sensitivity of $33\;\upmu e/\!\sqrt{\mathrm{Hz}}$~\cite{najera2024high}. Recent designs of charge qubits~(Cooper-pair boxes, CPBs) have achieved $10^{-7}\;e/\!\sqrt{\mathrm{Hz}}$ level by coupling a CPB to a microwave resonator and exploiting its quantum nonlinearity~\cite{tanarom2025characteristic}. However, these quantum sensors require cryogenic operation and complex microwave control, posing significant challenges to in vivo and clinical use. Cold-atom interferometers exploit matter-wave coherence to achieve exceptional inertial sensing precision, with laboratory gravimeters reaching $10^{-9}\;g$ sensitivity~\cite{menoret2018gravity} and fundamental constant measurements achieving fractional precision at the $10^{-11}$ level. However, their applications are confined to fundamental physics and gravity measurements, with minimal relevance to biomedical sensing due to bulky ultrahigh-vacuum laser-cooling apparatus and lack of biological compatibility.

Second-generation quantum medical sensors introduce phase- and coherence-based measurement protocols into biologically compatible sensing platforms. In practice, however, coherence is only partially exploited: most implementations rely on short-lived, weakly controlled coherence within ensembles of effectively independent probes, yielding sensitivities that are ultimately bounded by the SQL. Across NMR, NV centers, SQUIDs, and optically pumped magnetometers, translational progress has therefore been driven primarily by engineering advances such as improved field control, sensor miniaturization, array scaling, and signal processing, rather than by active coherence preservation as a scalable metrological resource. As a result, the practical performance ceiling of second-generation sensors is set by deployability, stability, and system integration rather than by fundamental quantum bounds, leaving substantial metrological headroom that coherence alone cannot access.

\noindent\textbf{Third-Generation Quantum Sensor: Entanglement-Enhanced Metrology.}\\

\begin{figure}[htbp]
    \centering
    \includegraphics[width=\columnwidth]{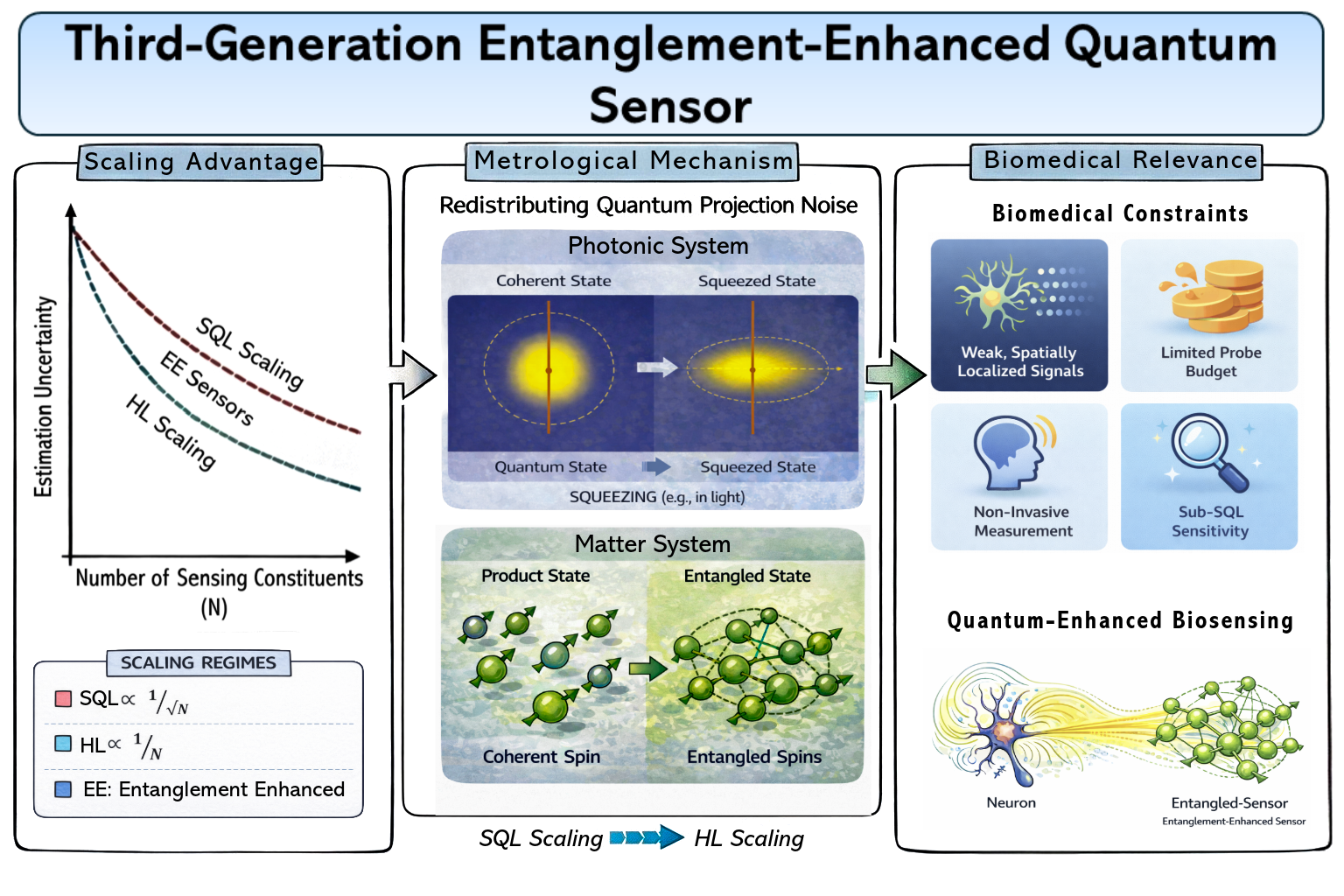}
    \caption{Entanglement-enhanced biomedical quantum sensing: the 3rd-generation advance.}
    \label{fig:3rd-gen}
\end{figure}

The third generation of quantum sensing is defined by the deliberate engineering of quantum correlations, including entanglement and spin squeezing, to enhance weak-signal estimation beyond classical limits. Unlike second-generation sensors, which rely on coherent evolution of individual quantum systems, third-generation architectures treat shared many-body correlations as the primary metrological resource, redistributing measurement noise across correlated probes to improve the detectability of weak biological signals without proportionally increasing biological burden. This transition is illustrated schematically in Fig.~\ref{fig:3rd-gen}.

This noise redistribution is particularly relevant for biomedical sensing, where signals such as neuronal magnetic fields, spin-labeled biomolecules, and radical-pair dynamics are intrinsically weak, spatially heterogeneous, and constrained by limits on repetition imposed by biological viability. Improving sensitivity through collective quantum correlations offers enhanced information extraction without increasing measurement backaction or exposure. Among the platforms under active investigation, nitrogen-vacancy centers in diamond and optomechanical systems receive particular attention for their potential room-temperature, biocompatible operation.

The third-generation module in Table~\ref{tab:merged_quantum_sensors} summarizes representative platforms spanning solid-state, atomic, photonic, and superconducting implementations, highlighting the diversity of entanglement resources, measurable quantities, reported sensitivity gains, and sensing distances. While several platforms already operate at or near room temperature, others require cryogenic or ultrahigh-vacuum environments, reflecting differing maturity levels. We briefly discuss representative advances across these platforms.

In nitrogen-vacancy center systems, recent experimental advances have established multiple entanglement-enabled pathways toward improved sensitivity under ambient conditions. Entanglement between an NV electronic spin and a proximal $^{13}$C nuclear spin enabled phase estimation beyond uncorrelated limits at room temperature~\cite{Liu2015}. These concepts were extended to nanoscale sensing by entangling closely spaced NV pairs, leading to enhanced magnetic sensitivity, improved spatial resolution, and common-mode noise rejection when probing external dark spins near the diamond surface~\cite{Zhou2025}. At the multi-qubit level, dipolar coupling between pairs of NV centers has been used to deterministically prepare two-qubit Bell states for entanglement-enhanced covariance magnetometry, reducing the readout-noise penalty from quadratic to linear and establishing entanglement as a practical nanoscale sensing resource~\cite{Rovny2025}. Related analyses have examined robustness under realistic noise models, showing that moderately entangled states can retain sensitivity gains over uncorrelated probes even in the presence of Markovian decoherence~\cite{PhysRevA.111.042605}. Schemes exploiting double-quantum transitions, globally addressed microwave control, and qutrit encodings further leverage the full spin-1 Hilbert space of the NV center to accelerate phase accumulation and suppress common-mode noise~\cite{morillas2025entanglement,Gassab_2025}. Ensemble-scale demonstrations of dipolar-interaction-mediated spin squeezing have additionally shown that correlated many-body states can be generated \emph{in situ} in disordered solids by selectively mitigating strongly coupled spin dimers~\cite{Wu2025}. Together, these results establish NV centers as a strong solid-state platform for third-generation biomedical sensing, enabling high-sensitivity, minimally invasive interrogation of spin-based biomarkers and dynamic biochemical processes.

Optomechanical platforms provide a distinct third-generation architecture by coupling optical fields to mechanical motion through radiation pressure, enhancing force and displacement sensitivity while reducing technical noise under ambient conditions. Cascaded configurations employing entangled mechanical modes aim to improve force detection and suppress common-mode disturbances~\cite{Xia2023}. Injection of optical squeezed light into microdisk resonators has enabled magnetometry below the shot-noise limit at room temperature~\cite{Li:18}. Direct measurements of quantum backaction correlations have enabled quantum-referenced thermometry without cryogenic cooling~\cite{doi:10.1126/science.aag1407}, while truncated nonlinear interferometry integrated with atomic force microscopy has demonstrated femtometer-scale displacement sensitivity~\cite{PhysRevLett.124.230504}. These advances suggest that optomechanical systems may suit correlation-enhanced sensing of force, displacement, temperature, and magnetic fields in biological environments, though scaling to clinical array formats remains an open challenge.

Trapped-ion systems, while operating under ultrahigh-vacuum conditions, provide a well-established laboratory platform for entanglement-enhanced sensing with long coherence times and high-fidelity quantum control. Early demonstrations established Heisenberg-limited phase scaling using small GHZ states~\cite{doi:10.1126/science.1097576}, subsequent experiments generated multipartite entangled states of more than ten ions with super-resolved phase evolution~\cite{PhysRevLett.106.130506}, and deterministic spin squeezing has produced several decibels of metrological gain~\cite{doi:10.1126/science.aad9958}. Although not directly deployable, the techniques developed in trapped-ion systems are expected to inform design of more practical platforms.

At the macroscopic scale, atomic vapor systems, particularly those operating in the spin-exchange relaxation-free regime, represent room-temperature platforms for entanglement-enhanced magnetometry with direct relevance to clinical neuroimaging and cardiography. Quantum non-demolition measurements in dense alkali vapors have produced singlet-type correlations among more than $10^{13}$ atoms, generating measurable spin squeezing and spatially extended entanglement~\cite{Kong2020}. Complementary approaches based on stroboscopic back-action-evading measurements have prepared spin-squeezed states in cesium vapor magnetometers, achieving several decibels of squeezing and improving magnetic induction tomography sensitivity beyond classical limits~\cite{PhysRevLett.130.203602}. Entanglement-assisted magnetic induction tomography has further demonstrated substantial noise reduction and improved spatial resolution, highlighting the potential of vapor-based systems for imaging weakly conductive biological tissue~\cite{PhysRevLett.130.203602}. Gradiometric configurations employing entangled twin beams have enabled simultaneous suppression of ambient magnetic interference and photon shot noise, achieving femtotesla-scale gradient sensitivity in unshielded environments~\cite{doi:10.1126/sciadv.adg1760}. For clinical applications such as magnetoencephalography, fetal cardiography, and noninvasive conductivity imaging, vapor-based systems are plausible near-term candidates and natural nodes for the local entangled arrays that constitute the first stage of the fourth-generation network roadmap.

Cold-atom Rydberg systems provide another atomic architecture for third-generation quantum sensing, leveraging large electric dipole moments and tunable long-range interactions. Initial demonstrations using these ensembles have achieved microwave electrometry approaching the standard quantum limit, achieving nanovolt-per-centimeter sensitivities over megahertz bandwidths~\cite{doi:10.1126/sciadv.ads0683}. Although these initial demonstrations relied primarily on coherent control rather than engineered entanglement, subsequent experiments have realized two-atom entanglement near nanophotonic surfaces, enabling micrometer-scale electric field gradient sensing within decoherence-free subspaces~\cite{PhysRevLett.132.113601}. Complementary theoretical proposals based on Rydberg dressing predict the generation of spin-squeezed states in extended two-dimensional atomic arrays through engineered long-range interactions~\cite{PhysRevLett.131.063401}. Rydberg interaction arrays have also demonstrated several-decibel spin squeezing in two-dimensional geometries, underscoring the metrological potential of engineered long-range interactions~\cite{Bornet2023}. Such array-based architectures naturally support spatially multiplexed sensing and suggest a scalable route toward distributed quantum sensor networks capable of imaging weak electric and magnetic field gradients in biological tissues.

Photonic interferometer architectures exploit engineered optical correlations, including squeezing, twin-beam intensity correlations, and entangled photon pairs, to redistribute measurement noise across correlated modes, improving information extraction per photon while remaining compatible with biologically constrained illumination. Quantum-enhanced Raman microscopy and nonlinear interferometric imaging have demonstrated sub-shot-noise sensitivity for chemical contrast and live-cell imaging~\cite{Casacio2021,Terrasson24}. Interferometric phase-imaging with entangled photons has achieved scan-free wide-field sensitivity improvements for protein microarray detection~\cite{Camphausen2021}, while entangled-photon interferometry has enabled adaptive aberration correction in scattering media~\cite{Cameron2024}. Quantum optical coherence tomography leverages Hong-Ou-Mandel interference for dispersion cancellation and enhanced axial resolution in multilayer biological samples, with recent phase-dependent protocols suppressing cross-reflection artifacts~\cite{YepizGraciano2022,li2025phase}. These photonic approaches enhance sensitivity and imaging fidelity without increasing illumination intensity, a critical constraint where photodamage must be minimized.

In contrast to room-temperature solid-state and optical approaches, superconducting architectures, including qubits, resonators, and parametric amplifiers, provide a cryogenic implementation of third-generation quantum sensing. Their strong nonlinearities and precise quantum control enable the preparation and manipulation of nonclassical states, including squeezing and multipartite entanglement, for enhanced measurement precision. Experiments have demonstrated criticality-enhanced force sensing by operating Josephson parametric oscillators near bifurcation points, achieving near-quantum-limited detection of sub-attonewton forces with potential relevance for ultrasensitive mechanical biosensing~\cite{2025PRXQ....6b0301B}. In the microwave domain, squeezed vacuum fields generated by parametric amplifiers have enhanced sensitivity in magnetic resonance spectroscopy, enabling improved detection of weak spin ensembles relevant to biochemical and materials systems~\cite{PhysRevX.7.041011}. Quantum illumination protocols using entangled microwave-optical photon pairs have further demonstrated sensitivity advantages for detecting low-reflectivity targets in thermal environments, suggesting potential pathways toward noninvasive biomedical imaging~\cite{doi:10.1126/sciadv.abb0451}. These developments illustrate how cryogenic superconducting platforms can leverage engineered quantum correlations for high-precision sensing, while also highlighting the practical trade-offs associated with low-temperature operation in biomedical contexts.

Third-generation quantum sensors demonstrate that engineered quantum correlations can now be prepared and characterized across a range of platforms operating under increasingly practical conditions, including room-temperature and ambient environments. Entanglement-enhanced sensitivity has been demonstrated in controlled laboratory settings, but the extent to which these scaling advantages persist in biologically complex systems remains uncertain and depends on decoherence, geometry, and integration constraints. Current implementations therefore represent a significant advance in correlation engineering rather than a completed route to clinical performance gains. In many platforms, entangled probe preparation, adaptive control, and hardware-level optimization are already embedded in the sensing protocol itself. This convergence of state engineering and protocol design points toward a fourth generation in which sensing, processing, and inference would operate within a single quantum architecture, bypassing the classical readout stage that currently separates measurement from analysis.

\noindent\textbf{Fourth Generation Quantum Medical Sensor: Quantum Learning.}\\

\begin{figure}[htbp]
    \centering
    \includegraphics[width=0.5\textwidth]{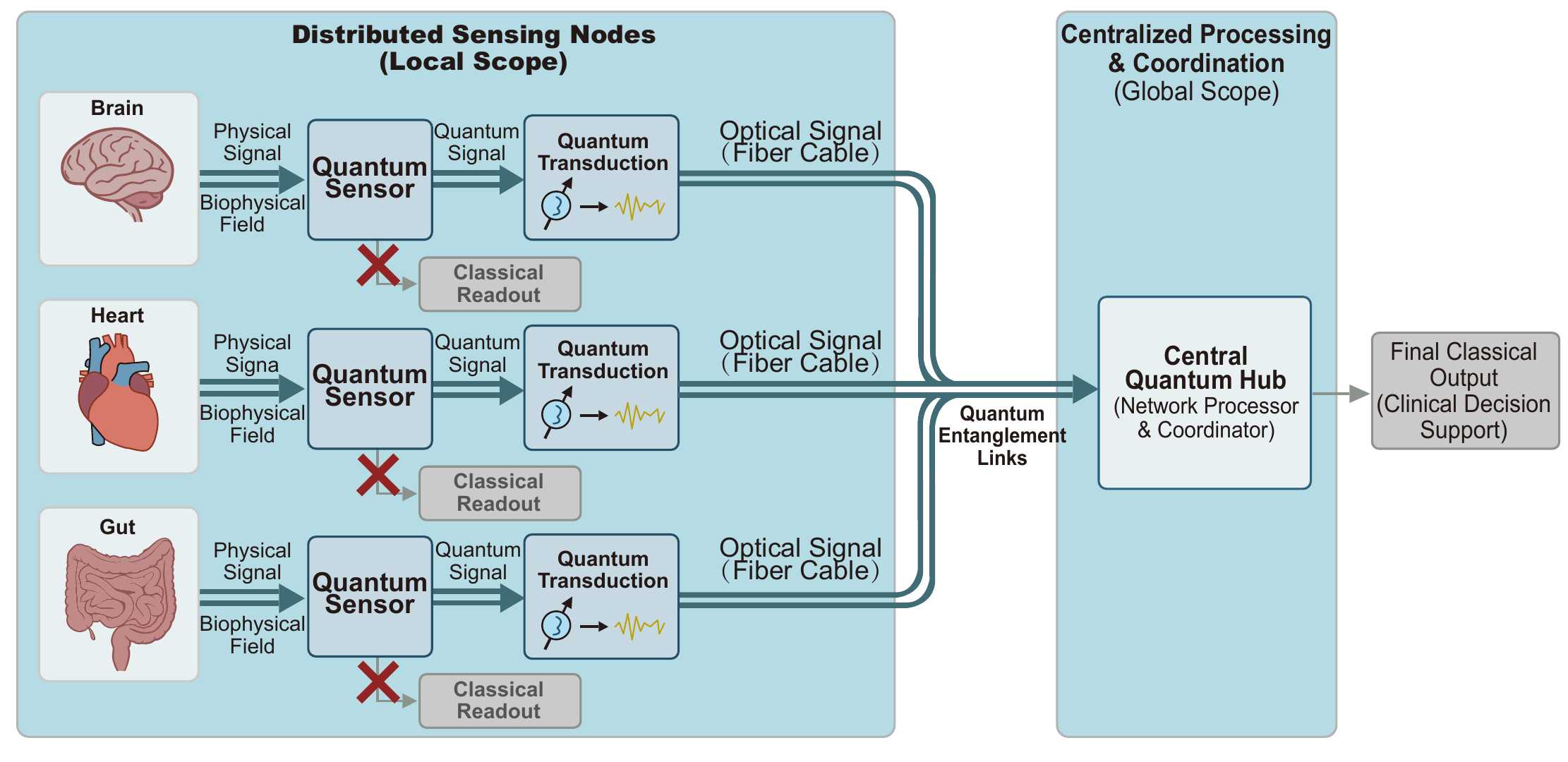}
    \caption{
    System architecture of the \textbf{fourth-generation quantum medical sensor}.
    Quantum signals originating from biological targets are captured by quantum sensors 
    (e.g., NV centers), transmitted via optical quantum transduction, and processed either 
    locally or within a coordinated sensor network. 
    The quantum processor performs quantum learning or inference tasks before producing 
    classical outputs for downstream medical analysis. 
    The dashed components indicate optional centralized coordination, which is disabled in 
    fully distributed sensor-network operation where each node performs independent quantum 
    processing and participates in entanglement-based sensing protocols.
    }
    \label{fig:fourth_generation_quantum_medical_sensor}
\end{figure}
Fourth-generation quantum medical sensors represent a qualitative departure from earlier biomedical sensing architectures by eliminating the classical measurement bottleneck. Earlier quantum sensors ultimately collapse quantum states into classical readouts at an intermediate stage, discarding quantum correlations before downstream processing. In contrast, fourth-generation systems are defined by the end-to-end integration of quantum sensing with quantum computation and quantum learning: sensors forward quantum states directly to downstream quantum processors, enabling coherent processing without intermediate classical readout, and supporting distributed sensing networks~\cite{zhang2021distributed,zhao2021field}. Two boundaries delimit this generation. First, where machine learning has entered biomedical sensing discussions as classical post-processing of measurement records, fourth-generation systems perform inference in the quantum domain before readout, which makes learning-theoretic separations accessible. Second, where quantum computing has been positioned as a longer-horizon healthcare technology separate from near-term sensing~\cite{LoFiego2025}, the fourth generation identifies sensing-computation convergence, enabled by quantum transduction, as the trajectory. In the remainder of this section, we first examine the quantum learning advantages that underpin this generation, then describe the network architectures that enable multi-node sensing, and finally discuss the role of quantum transduction in bridging heterogeneous platforms and enabling coherent quantum information transfer across nodes.

\emph{Quantum learning advantage.} Fourth-generation quantum medical sensors derive a central part of their advantage from the integration of quantum learning directly into the sensing pipeline, enabling information to be extracted with reduced sample complexity and enhanced precision. This capability is particularly critical in biomedical contexts, where repeated probing may disrupt biological function, cause tissue damage, or expose patients to harmful stimulation. Recent theoretical and experimental work has established provable quantum advantages in learning from experiments, demonstrating reductions in both the number of required quantum state copies and the total experimental time~\cite{huang2023learning,huang2022quantum,oh2024entanglement,liu2025quantum}.


A prominent application of these advantages is learning the dynamics of unknown systems, typically parameterized as $H(\theta) = \sum_j \theta_j H_j$. By leveraging coherent phase accumulation and entanglement-assisted measurements, quantum protocols can achieve Heisenberg-limited scaling, requiring only $\mathcal{O}(\varepsilon^{-1})$ total evolution time to reach precision $\varepsilon$, strictly outperforming the $\mathcal{O}(\varepsilon^{-2})$ scaling inherent to classical strategies~\cite{huang2023learning}. Furthermore, under suitable access models, quantum methods exponentially reduce the number of distinct experiments required, scaling only polylogarithmically with the target precision~\cite{huang2023learning,huang2022quantum}. These theoretical gains in sampling efficiency are clinically meaningful in biomedical settings, where minimizing excessive probing is critical to avoiding tissue disruption or harmful radiation exposure. The classical shadow framework further enables efficient simultaneous prediction of many observables from a small number of quantum state copies~\cite{huang2020predicting,huang2022quantum}. Instead of fully reconstructing the sensor-generated quantum state, the protocol applies random unitaries followed by computational-basis measurements to produce classical shadows from which a large family of observables can be estimated. For a set $\mathcal{F} = \{O_j\}$ of sensing-relevant observables, the number of state copies required to estimate all $y_j(u) = \mathrm{Tr}(O_j\,\rho(u))$ within additive error $\varepsilon$ with high probability scales as
\begin{equation}
    N_{\mathrm{copies}} = \mathcal{O}\!\left(\frac{\max_j \mathrm{Var}[\hat{y}_j]\cdot\log|\mathcal{F}|}{\varepsilon^2}\right),
\end{equation}
yielding a logarithmic dependence on the number of target observables $|\mathcal{F}|$, in contrast to the linear scaling of independent estimation or the exponential cost of full state tomography. This logarithmic scaling might in principle enable multimodal biomedical monitoring, where a single measurement campaign simultaneously characterizes multiple field components, spectral features, or spatial modes under tight acquisition-time constraints. A further fundamental advantage arises from quantum memory: learners that can coherently store and jointly process multiple state copies before choosing which measurements to perform are provably more efficient than classical learners restricted to measurement transcripts. For certain observable families, this separation is exponential in the number of required experimental interactions~\cite{huang2022quantum}.

Quantum-enhanced hypothesis testing leverages entangled states and coherent measurements to discriminate between candidate physical models or detect weak signals with fewer experimental resources than any classical strategy permits~\cite{oh2024entanglement,liu2025quantum}. When quantum sensors are connected to processors capable of joint entangling operations, the resulting measurement statistics might resolve binary or multi-class hypotheses, such as the presence of a biomarker or the classification of tissue pathology, with error probabilities that decrease exponentially faster per sample than classical bounds allow. Crucially, these learning-theoretic improvements are not mutually exclusive with traditional metrological gains but can be compounded within a single sensing protocol, potentially enabling diagnostic workflows that are simultaneously more precise, less invasive, and more informative than existing classical or coherence-only approaches. While these sample-complexity results are established in quantum information theory, a central aim of this Perspective is their clinical translation: reduced probing as reduced tissue exposure and dose, shadow-tomography scaling as multimodal monitoring under acquisition-time constraints, hypothesis-testing error exponents as diagnostic classification performance, and optimization latency measured against intraoperative timing budgets.

The theoretical advantages described above rely on coherent multi-copy processing or entanglement-assisted measurements that presuppose fault-tolerant quantum hardware beyond current capabilities. As a complementary near-term strategy, variational quantum circuits (VQCs) pursue a fundamentally different form of quantum learning: rather than implementing provably optimal protocols, VQCs use parameterized circuits optimized via hybrid quantum-classical loops to maximize task-specific objectives such as the quantum Fisher information~\cite{arunkumar2023quantum,zhuang2019physical,srivastava2025variational}. Although this heuristic approach sacrifices provable optimality for hardware compatibility, early demonstrations on NV center ensembles and superconducting platforms suggest that meaningful sensing enhancements are achievable within current noise and connectivity constraints. Complementary approaches, such as the incorporation of quantum logic operations and assisted quantum memory in solid-state platforms, have further demonstrated suppression of readout noise, addressing one of the primary bottlenecks in spin-based biomedical sensing.

Despite these advances, practical deployment faces significant latency challenges. VQCs require iterative hybrid optimization loops in which a classical co-processor repeatedly updates circuit parameters based on measurement outcomes, with end-to-end latency per optimization step reaching milliseconds to seconds depending on circuit depth, parameter count, and shot budget. For biomedical applications demanding real-time feedback, such as intraoperative neural monitoring or closed-loop neurostimulation, these timescales may exceed clinically relevant thresholds. Whether near-term quantum processors can satisfy such timing constraints while maintaining sufficient circuit expressibility and noise resilience remains an open translational challenge.

\emph{Quantum sensor networks.} A quantum sensor network jointly operates multiple quantum sensors while preserving quantum resources across nodes. Only networks that maintain quantum coherence or entanglement qualify as genuine quantum sensor networks; those that merely aggregate classical readouts cannot sustain quantum advantage after measurement-induced collapse. These networks divide into two regimes: local and distributed, each addressing different biomedical constraints.
\begin{figure}[htbp!]
    \centering
    \includegraphics[width=0.5\textwidth]{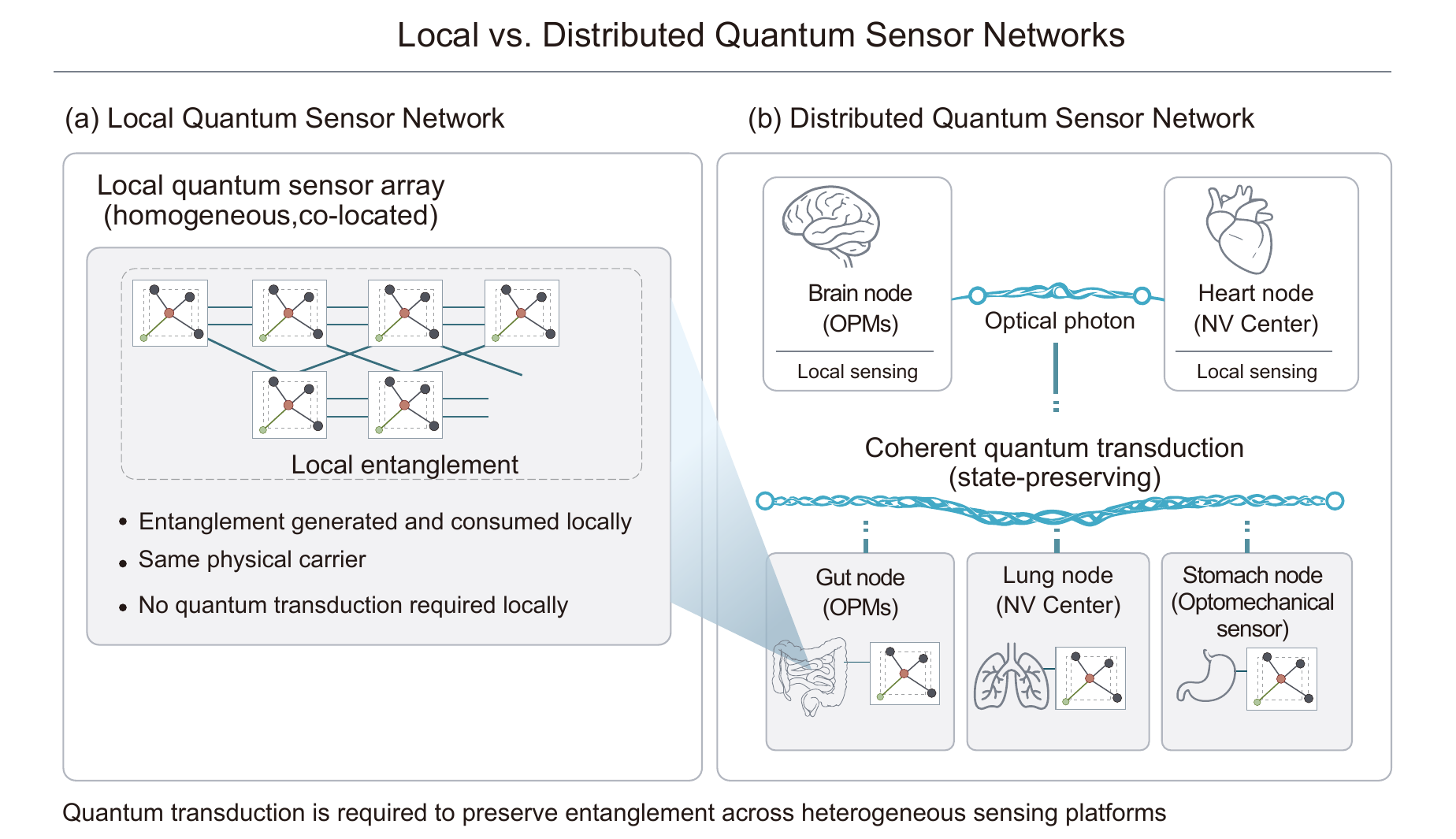}
    \caption{
    Conceptual comparison between a \textbf{local quantum sensor network} and a 
    \textbf{distributed quantum sensor network}. 
    In the local configuration, entanglement is generated and consumed entirely within a 
    co-located sensor array, eliminating the need for quantum transduction. 
    In contrast, the distributed sensor network connects spatially separated sensing nodes 
    (e.g., brain, cardiac, or gastric sensors) via quantum transduction, enabling long-range 
    quantum correlations across heterogeneous biological sites.
    }
    \label{fig:local_vs_distributed_quantum_network}
\end{figure}
Local quantum sensor networks consist of co-located homogeneous sensors (e.g., NV center ensembles or arrays within a single organ) that can be entangled locally for enhanced sensitivity and noise suppression through collective quantum effects. Because all nodes operate on the same physical carrier and are spatially proximate, quantum states are generated, manipulated, and consumed entirely within a single platform, enabling entanglement-assisted sensing without inter-platform conversion or quantum transduction. This regime is well suited for array-level biomedical measurements where multiple probes must jointly monitor a spatially extended target while remaining physically accessible for local entangling operations.

Distributed quantum sensor networks arise when sensors are spatially separated or physically heterogeneous, as occurs when probes span distinct organs or clinical sites. Different sensor types (e.g., NV centers, optomechanical transducers) encode quantum information in incompatible physical carriers, forcing current protocols to measure at each node individually and aggregate classical data. This intermediate readout irreversibly destroys coherence, entanglement, and high-dimensional Hilbert-space correlations~\cite{holevo1973bounds,aaronson2016complexity}, projecting the exponentially rich state of an $n$-qubit system into a low-dimensional classical record. The premature-measurement problem is therefore the central bottleneck separating classical sensor fusion from genuine quantum-enhanced distributed sensing.

\emph{Quantum transduction.} Quantum transduction addresses the premature-measurement bottleneck identified above by coherently converting quantum states between heterogeneous physical carriers without intermediate classical readout~\cite{zhong2022quantum,lauk2020perspectives}. Unlike classical transducers that convert analog signals between physical domains, quantum transduction must preserve the full quantum state including superposition and entanglement, imposing stringent constraints on conversion fidelity, added noise, and bandwidth. By converting quantum states, rather than classical signals, into a common carrier (typically optical photons at telecom wavelengths for long-distance transmission), transduction ensures that sensor-generated quantum resources remain available for joint processing across nodes~\cite{lauk2020perspectives,zhong2022quantum}. The resulting distributed architecture supports \emph{joint inference}: simultaneous extraction of correlated information from anatomically separated sensors that is inaccessible through independent classical measurements. Entanglement-enhanced precision enables distributed protocols to surpass the standard quantum limit: for global parameters, an $M$-node entangled network achieves precision scaling as $1/M$ rather than the $1/\sqrt{M}$ of classically fused independent records~\cite{zhang2021distributed,ge2018distributed,eldredge2018optimal}. More broadly, a network of $M$ entangled nodes with $n$ qubits each supports joint states whose classical description requires exponentially many parameters~\cite{lloyd2021quantum}; although the Holevo bound caps the classical information extractable from such a state at $Mn$ bits per copy~\cite{holevo1973bounds}, coherent joint processing can exploit these correlations \emph{before} readout, enabling exponential sample-complexity separations for suitable learning tasks~\cite{huang2022quantum}. Quantum transduction is therefore not merely a peripheral engineering challenge but a necessary architectural component that delineates the boundary between third- and fourth-generation quantum medical sensing systems.

Complementing institutional roadmaps for quantum biosensing~\cite{LoFiego2025}, the staged progression outlined here specifies physical rather than organizational milestones. The progression from current demonstrations to fully distributed quantum medical sensor networks will proceed through several intermediate stages. In the \emph{near term} (approximately 2025-2030), the most accessible implementations are local entangled sensor arrays, such as NV center ensembles or co-located OPM clusters, that exploit intra-platform entanglement for enhanced sensitivity without requiring quantum transduction or long-range coherence. In the \emph{mid term} (approximately 2030-2040), advances in transduction fidelity and quantum memory lifetimes are expected to enable point-to-point quantum links between a small number of heterogeneous sensor nodes within a single clinical suite~\cite{zhong2022quantum}. In the \emph{long term} (beyond 2040), contingent on fault-tolerant quantum computation and scalable quantum networking, distributed quantum data centers could provide the computational backbone for population-scale medical sensing, hosting quantum processors, quantum random-access memory (QRAM) with $\mathcal{O}(\log N)$-time access to quantum data entries~\cite{liu2023data,liu2024quantum}, and network interfaces that enable coherent aggregation of sensor data from geographically distributed clinical facilities.

\section{Cross-Cutting Considerations}
\subsubsection*{Bandwidth Matching and Neural Signal Hierarchy}

\begin{table*}[htbp!]
\caption{Bandwidth matching between neural signal hierarchy and quantum sensing
platforms. The header row shows each platform's usable bandwidth; suitability
is indicated as: \checkmark~(well matched), $\circ$~(marginal), $\times$~(mismatched).
Frequency entries denote the bandwidth a sensor must sustain to reproduce the signal
without temporal distortion, which for the fastest signals exceeds the nominal signal
repetition rate.}
\label{tab:bandwidth_matching}
\centering
\setlength{\tabcolsep}{8pt}
\renewcommand{\arraystretch}{1.3}
\resizebox{\textwidth}{!}{%
\begin{tabular}{lcccc}
\toprule
\textbf{Signal Type} & \textbf{Frequency / req.\ bandwidth} & \textbf{SQUID} & \textbf{OPM} & \textbf{NV Center} \\
\midrule
\textit{Sensor bandwidth} & 
& DC-$\sim$1\,kHz (flux-locked loop)~\cite{clarke2006squid,baillet2017magnetoencephalography}
& DC-$\sim$150\,Hz (SERF)$^{\dagger}$~\cite{shah2013compact,boto2018moving}
& DC-kHz-GHz~\cite{barry2020sensitivity,herb2025quantum,fortman2021electron} \\
\midrule
Cortical rhythms ($\alpha$, $\beta$, $\gamma$)~\cite{whittington2018future} 
  & $<$100\,Hz      & \checkmark & \checkmark & \checkmark \\
High-gamma oscillations                    
  & 80-200\,Hz     & \checkmark & $\circ$    & \checkmark \\
Evoked responses (MEG/MCG)                 
  & 1-100\,Hz      & \checkmark & \checkmark & \checkmark \\
Peripheral nerve CAPs~\cite{pelot2017modeling}     
  & 100\,Hz-5\,kHz & $\circ$    & $\times$   & \checkmark \\
Single-neuron action potentials            
  & $\sim$1\,kHz    & $\times$   & $\times$   & \checkmark \\
Fast invasive LFP                          
  & 1-8\,kHz       & $\times$   & $\times$   & \checkmark \\
Ion-channel gating (fastest transitions)$^{\ddagger}$~\cite{Sigg2003,Hartel2019}
  & 10-100\,kHz    & $\times$   & $\times$   & $\circ$    \\
Molecular processes (MHz)~\cite{doi:10.1073/pnas.1001832107} 
  & $>$1\,MHz       & $\times$   & $\times$   & \checkmark~\cite{lovchinsky2016nuclear} \\
\bottomrule
\multicolumn{5}{l}{\footnotesize\textit{Overqualified}: NV bandwidth exceeds signal requirements; sensitivity may be the limiting factor.}\\
\multicolumn{5}{l}{\footnotesize$^{\dagger}$3-dB bandwidth of SERF-mode OPMs (${\sim}$100-150\,Hz); non-SERF total-field OPMs reach ${>}100$\,kHz at reduced sensitivity~\cite{jaufenthaler2021pulsed}.}\\
\multicolumn{5}{l}{\footnotesize$^{\ddagger}$Required recording bandwidth set by the 3-12\,$\upmu$s rise times of the fastest resolved gating transitions, not a signal repetition rate.}
\end{tabular}%
}
\end{table*}
The generational classification introduced above describes biomedical quantum sensors according to the depth of quantum resources employed, ranging from energy-level readout in first-generation devices to entanglement-enhanced metrology and quantum learning in later generations. However, the suitability of a sensing platform for a specific biomedical application depends not only on quantum resource depth but also on several orthogonal physical parameters. Bandwidth, defined as the frequency range over which a sensor maintains useful sensitivity, emerges as a fundamental and often underappreciated dimension. Unlike sensitivity, which can often be improved through engineering strategies such as averaging or array scaling, bandwidth is constrained primarily by the underlying sensing mechanism and operating mode. As a result, bandwidth and generational classification form two independent axes: sensors of the same quantum generation may exhibit very different usable bandwidths depending on their physical implementation.

From a neurophysiological perspective, the fastest canonical electrical signal produced by neurons is the action potential, with a characteristic duration of approximately 1~ms~($1~\text{kHz}$), a limit set by ion-channel refractory dynamics rather than engineering constraints. Resolving individual neuronal spikes or microcircuit-level dynamics therefore requires sensing modalities capable of operating in the kilohertz regime and in close spatial proximity to the signal source. Experimentally recorded brain signals represent the superposition of many asynchronously firing neurons, producing a broad spectral distribution; recent invasive recordings have reported local field potential activity extending beyond 1~kHz and into the 2-8~kHz range in pathological conditions such as epilepsy~\cite{Usui2010,brazdil2023ultra}. At even smaller spatial scales, ion-channel gating is intrinsically multi-scale: single-channel dwell times run from seconds and milliseconds down to ${\sim}10\;\upmu\mathrm{s}$ flicker closures, and gating-charge transitions as fast as $3$-$12\;\upmu\mathrm{s}$ have been resolved~\cite{hille1978ionic,Sigg2003,Shapovalov2004}. Resolving these fastest transitions, rather than merely detecting gating, sets the sensing requirement at a $10$-$100\;\mathrm{kHz}$ recording bandwidth~\cite{Hartel2019}.

Table~\ref{tab:bandwidth_matching} summarizes this temporal hierarchy against the characteristic bandwidths of representative quantum sensing platforms.

OPMs, particularly those operating in the SERF regime, achieve femtotesla-level sensitivity but typically exhibit 3-dB bandwidths of approximately 100-150~Hz (DC-${\sim}150$~Hz)~\cite{shah2013compact,boto2018moving}, a limit set by the underlying spin-relaxation physics; non-SERF total-field OPMs achieve bandwidths exceeding 100~kHz~\cite{jaufenthaler2021pulsed}, but at substantially lower sensitivity. This bandwidth is well matched to conventional MEG, whose signal content lies primarily below a few hundred hertz~\cite{baillet2017magnetoencephalography}, but fundamentally restricts access to faster neural dynamics such as spike-level or microcircuit activity. Moreover, current OPM systems are designed for macroscopic, far-field measurements with sensor-source distances on the order of centimeters, inherently averaging over large neural populations and suppressing spatially localized or high-frequency components.

In contrast, solid-state NV centers in diamond inherently support broadband magnetic sensing owing to their fast spin dynamics and room-temperature operation~\cite{barry2020sensitivity,fortman2021electron}. In near-field geometries, where sensor-source distances can be reduced to the micrometer-to-millimeter scale, magnetic fields scale as $1/r^{3}$, enabling NV sensors to access fast, localized neural signals in the kHz regime and beyond. However, near-field operation currently limits NV magnetometry to surface-accessible or surgically exposed tissue, and single-sensor sensitivity in the picotesla regime remains insufficient for detecting deep cortical sources at macroscopic standoff distances.

\begin{table*}[htbp!]
\caption{Clinical needs versus quantum sensor capability matching across representative biomedical applications.
Each row identifies a specific unmet clinical need, the core limitations of existing devices,
the most suitable quantum sensing platform (with generational classification),
the physical origin of the quantum advantage,
and the key translational bottleneck. TRL (Technology Readiness Level) values follow the NASA/Mankins scale  and refer specifically to the developmental stage of the Best-Matched Quantum Platform in the context of the identified clinical need, ranging from basic principles (TRL 1) to systems proven in operational deployment (TRL 9)~\cite{mankins2009technology}.}
\label{tab:matching}
\centering
\scriptsize
\begin{tblr}{
  colspec = {p{1.4cm} p{2.0cm} p{2.7cm} p{2.1cm} p{1.9cm} p{2.6cm} p{0.8cm} p{2.1cm}},
  row{1} = {bg=headerblue, fg=white, font=\bfseries\scriptsize, valign=m},
  row{2} = {bg=catblue, font=\bfseries\scriptsize},
  row{7} = {bg=catblue, font=\bfseries\scriptsize},
  row{11} = {bg=catblue, font=\bfseries\scriptsize},
  row{14} = {bg=catblue, font=\bfseries\scriptsize},
  row{17} = {bg=catblue, font=\bfseries\scriptsize},
  row{3,5,8,10,13,16} = {bg=rowalt},
  cell{2}{1} = {c=8}{l}, cell{7}{1} = {c=8}{l},
  cell{11}{1} = {c=8}{l}, cell{14}{1} = {c=8}{l},
  cell{17}{1} = {c=8}{l},
  rowsep = 1.5pt, colsep = 2.5pt,
  hline{1,Z} = {0.08em},
  hline{2} = {0.05em},
}
Clinical Domain & Unmet Clinical Need & Limitations of Existing Devices & Requirement Parameters & Best-Matched Quantum Platform (Gen.) & Origin of Quantum Advantage & TRL & Key Translational Bottleneck \\
A.~Neuroscience & & & & & & & \\
Whole-brain functional imaging
& ms-resolved whole-brain imaging during natural movement
& fMRI: BOLD delay ${\sim}\,5$\,s, immobility; SQUID-MEG: rigid helmet, ${\sim}\,200$ units globally; EEG: localization error up to 25\,mm
& ${<}\,50\;\mathrm{fT}/\!\sqrt{\mathrm{Hz}}$; ${<}\,1$\,ms; ${<}\,5$\,mm; non-ionizing
& OPM-MEG (2nd Gen.) wearable array
& Cryogen-free $\to$ wearable $\to$ 3-$5\times$ signal gain; 15-sensor OPM $\approx$ 62-ch SQUID
& 5-7
& MSR still required; sensor scalability; insufficient clinical validation \\
Single-neuron electrophysiology
& Non-invasive single-neuron AP detection
& Microelectrodes: invasive; fMRI: ${\sim}\,1$\,Hz; MEG/EEG: cannot resolve single cells
& ${\sim}$\,nT (axon surface); ${\sim}\,30\;\upmu$s; ${<}\,100\;\upmu$m; near-field
& NV-diamond magnetometer (2nd Gen.)
& $\upmu$m standoff $\to$ near-field gain $\to$ nT-scale single-AP fields at the axon surface; NV Center kHz-GHz bandwidth
& 3-5
& 532\,nm penetration 6.5\%/mm; surface only; sensitivity gap 1-2 orders \\
Pre-surgical epilepsy localization
& High-precision seizure-focus mapping reducing reliance on invasive iEEG
& iEEG: craniotomy, high risk; SQUID-MEG: inaccessible; EEG: poor localization
& ${<}\,20\;\mathrm{fT}/\!\sqrt{\mathrm{Hz}}$; ${<}\,5$\,ms; ${<}\,3$\,mm; non-invasive
& OPM-MEG (2nd Gen.) high-density array
& On-scalp $\to$ higher sampling; ictal recording feasible; He-4 OPM validated
& 6-7
& 510(k) pathway clear; large-scale trials needed; EMI management \\
Early neurodegenerative detection
& Pre-symptomatic biomarkers for AD/PD
& MRI: structural lag; PET-A$\beta$: radiation, costly; EEG: insufficient sensitivity
& Subtle rhythm changes; ms-scale; safe for repeated monitoring
& OPM-MEG (2nd) + quantum learning (4th Gen.)
& $\alpha$/$\beta$ rhythm detection; wearable $\to$ longitudinal; Gen~4: adaptive inference
& 4-6
& Biomarker validation lacking; cohort studies needed; data standardization \\
B.~Oncology & & & & & & & \\
Early cancer metabolic detection
& Detect metabolic abnormalities before morphological changes
& PET-FDG: 4-5\,mm, low Stage~I; ctDNA: 7\% Stage~I lung; CT/MRI: phenotypic
& Single-cell metabolism; ${<}\,1$\,mm; no radiation; repeatable
& Hyperpol.\ $^{13}$C MRI (2nd Gen.)
& ${>}\,10{,}000\times$ enhancement $\to$ real-time metabolic imaging; Warburg; no radiation
& 5-6
& Short $T_{1}{\sim}\,60$\,s; requires DNP; cost and workflow \\
Molecular tumor imaging
& Specific labeling and micro-metastasis visualization
& Gd: non-specific, nephrotoxic; PET: cyclotron; fluorescence: photobleaching, shallow
& Molecular sensitivity; target-specific; low toxicity; ${>}\,1$\,cm penetration
& Quantum dots (1st Gen.) NIR fluorescence
& Size-tunable NIR $\to$ deep penetration; peptide $\to$ targeting
& 4-6
& Cd-QD toxicity; non-degradable; De Novo regulatory \\
Intraop.\ margin assessment
& Real-time tumor boundary ID during surgery
& Intraop MRI: slow; frozen section: 20-30\,min; fluorescence: limited specificity
& ${<}\,\text{100 }\upmu$m; real-time; non-destructive; surgery-compatible
& NV wide-field magnetic imaging (1st-2nd Gen.)
& Immunomagnetic labeling $\to$ NV wide-field; sub-cellular resolution
& 3-4
& Surface only; pre-labeling; workflow not validated \\
C.~Cardiology & & & & & & & \\
Inflammatory cardiomyopathy screening
& \quad Non-invasive rapid screening
& Echo: operator-dependent; cardiac MRI: slow, expensive; ECG: insensitive to deep lesions
& ${\sim}\,\mathrm{pT}$; real-time; multi-channel; non-invasive
& SQUID-MCG (2nd Gen.) $\to$ OPM-MCG
& SQUID-MCG: PPV ${\sim}\,93\%$; OPM: cryogen-free $\to$ portable; contactless
& SQUID 7-8; OPM 5-6
& SQUID: cost; OPM-MCG: validation needed; no reimbursement \\
Fetal cardiac monitoring
& Non-invasive fetal arrhythmia diagnosis
& Fetal ECG: maternal contamination; fetal echo: poor temporal resolution
& ${\sim}\,\mathrm{pT}$; through maternal tissue; absolute safety; ms-scale
& OPM array (2nd Gen.) unshielded gradiometer
& $B$-field unattenuated through tissue; undistorted fetal signal; wearable
& 5-6
& Maternal motion artifacts; interference; protocol standardization \\
D.~Molecular / Cellular Detection & & & & & & & \\
Intracellular label-free sensing
& Real-time $T$/pH/$B$ inside living cells
& Fluorescent probes: photobleaching; EM: fixed samples; optical: diffraction ${\sim}\,200$\,nm
& ${<}\,100$\,nm; $\upmu$s; non-toxic; quantitative
& Fluorescent nanodiamond (1st-2nd Gen.)
& Chemically inert; nanoscale thermometry (${\sim}\,\mathrm{mK}$); single-protein NMR
& 4-5
& Size control; sub-toxic stress; research-to-IVD path \\
Single-molecule biomarker detection
& Ultra-sensitive detection at ultra-low concentrations
& ELISA: LOD ${\sim}\,\mathrm{pM}$; mass spec: sample prep; PCR: nucleic acids only
& Single-molecule; specific; rapid (POC); affordable
& SET/QD (1st Gen.) + NV relaxometry (2nd Gen.)
& SET: single-charge; NV: wash-free immunoassay; QD-FRET: molecular distance
& 3-5
& SET: cryogenic operation or sub-10\,nm island fabrication for room-temperature devices; NV standardization; competing digital ELISA \\
E.~Distributed Multi-Organ Monitoring (Future) & & & & & & & \\
Multi-organ coordinated monitoring
& Simultaneous brain-heart-gut quantum-correlated monitoring
& Current multimodal: classical fusion only; no cross-organ quantum correlations
& Distributed; entanglement preserved; adaptive real-time inference
& Distributed QSN (4th Gen.)
& Quantum transduction across heterogeneous nodes; VQC optimization; Heisenberg $\Delta{\sim}\,1/N$
& 1-3
& Transduction efficiency low; biological decoherence; ${>}\,10$\,yr \\
\end{tblr}
\end{table*}

\subsubsection*{Bridging Clinical Bottlenecks and Technological Maturity}
The successful clinical translation of quantum sensing hinges on a strategic alignment between the depth of quantum resource utilization and the multifaceted demands of biological environments. As recent commentary has emphasized, absolute sensitivity is not the only clinically relevant metric: proximity to the target, biocompatibility, and robustness jointly determine clinical performance~\cite{LoFiego2025}. The tables below systematize this observation, tabulating quantum-resource depth against translational readiness device by device and showing that the two axes are uncorrelated across the current device landscape. As systematically mapped in Table~\ref{tab:matching}, established diagnostic modalities have largely reached performance ceilings dictated by classical physics and infrastructure limitations. Conventional MRI, for instance, is constrained by exceptionally low thermal spin polarization (${\sim}\,10^{-5}$), necessitating prolonged acquisition times and absolute patient immobility. Similarly, the reliance of SQUID-based MEG on cryogenic cooling (${\sim}\,4\;\mathrm{K}$) imposes rigid dewar geometries that force a significant sensor-to-source standoff distance, inherently suppressing weak biomagnetic signals. A more granular breakdown of these specific device bottlenecks and their physical origins is provided in Table~\ref{tab:dualaxis} (Appendix~\ref{appendix: specification}).

Quantum sensors address these unmet clinical needs by leveraging coherence and entanglement to redefine the boundaries of sensitivity and accessibility. Second-generation platforms, such as OPMs, represent a pivotal breakthrough by eliminating cryogenic requirements and enabling wearable, high-density arrays. This transition addresses the primary bottleneck of ``immobility'' in neuroimaging, offering a three- to fivefold signal gain through reduced standoff distance while permitting natural subject movement. For metabolic oncology, the integration of hyperpolarized $^{13}$C overcomes polarization limits to provide ${>}\,10{,}000\times$ signal enhancement~\cite{ArdenkjrLarsen2003,sushentsev2022hyperpolarised}, enabling real-time, radiation-free metabolic imaging where traditional PET remains limited by cyclotron dependence and ionizing exposure. The historical evolution of these sensitivity gains across different sensing modalities is summarized in Table~\ref{tab:bottleneck} (Appendix~\ref{appendix: specification}).

Within the multiscale landscape of biomedical sensing, NV centers in diamond are particularly suited to near-field, high-bandwidth applications requiring resolution of localized microcircuits. Crucially, these two modalities are not competitors but clinical complements: OPMs excel at whole-brain network monitoring and wearable neuroimaging, while NV sensors are uniquely positioned for near-field, high-bandwidth interrogation of fast neural dynamics at the single-neuron and microcircuit level, as detailed in the bandwidth hierarchy of Table~\ref{tab:bandwidth_matching}.

\section{Conclusion}

In this perspective, we introduced a generational framework that organizes the evolution of biomedical quantum sensors according to the depth of quantum resources employed, spanning energy-level readout in first-generation systems, coherence-based metrology in second-generation platforms, engineered entanglement and squeezing in third-generation architectures, and integrated quantum learning and information-processing strategies in a prospective fourth generation. This taxonomy provides conceptual clarity in a rapidly expanding field where technological maturity, physical constraints, and quantum resource utilization are often treated interchangeably.

First-generation platforms demonstrate that the presence of quantized energy levels alone does not guarantee quantum advantage. Second-generation sensors harness coherent phase evolution as a metrological resource and have achieved clinically meaningful deployment in technologies such as MRI, SQUID-based magnetometry, and optically pumped magnetometers, although many implementations remain limited by partial coherence utilization and ensemble averaging rather than fundamental quantum scaling. Third-generation architectures introduce a qualitative transition by deliberately engineering many-body correlations to redistribute measurement noise and enhance sensitivity; experimental demonstrations across solid-state, atomic, photonic, and superconducting systems confirm measurable gains under controlled laboratory conditions, even as translation to complex biological environments remains constrained by decoherence, scalability, and integration challenges. The proposed fourth generation extends this trajectory by embedding sensing within broader quantum information-processing architectures. By integrating adaptive control, variational optimization, distributed networks, and transduction interfaces, these systems aim to enable data-driven inference directly at the physical layer, particularly in biomedical contexts where safety and invasiveness limit measurement repetition. However, quantum processing and learning impose substantial computational demands: variational quantum circuits require iterative classical-quantum optimization loops whose cost scales with circuit depth, qubit count, and the number of variational parameters, while distributed quantum sensor networks add communication and synchronization overhead. Whether these computational requirements can be met within the real-time latency constraints of clinical decision-making, such as intraoperative guidance or continuous physiological monitoring, remains an open and important deployment challenge.

Beyond clinical translation, quantum sensors are opening new avenues in fundamental medical science that merit equal emphasis. At the subcellular scale, nitrogen-vacancy magnetometry has enabled direct observation of magnetic signatures from individual neurons, magnetotactic bacteria, and immunolabeled tumor sections, providing experimental access to biological phenomena that were previously below the detection threshold of any existing technique. In pharmacology, nanoscale quantum sensors are beginning to enable label-free, real-time monitoring of drug-target binding kinetics at the single-molecule level~\cite{lovchinsky2016nuclear}, complementing conventional high-throughput screening with mechanistic insight~\cite{santagati2024drug,blunt2022perspective}. Nanoscale nuclear magnetic resonance enabled by single NV centers has achieved chemical spectroscopy of zeptoliter sample volumes, establishing a new experimental paradigm for structural biology of membrane proteins, metabolites, and biomolecular complexes under native conditions. These contributions to basic biomedical research, spanning biophysics, neuroscience, oncology, and pharmacology, demonstrate that the impact of quantum sensing extends well beyond clinical diagnostics and should inform the prioritization of future research directions alongside clinical integration efforts.

Generational depth alone does not determine biomedical suitability. Bandwidth, sensor-sample geometry, thermal compatibility, and clinical safety impose independent constraints that shape practical deployment, and physiological processes span spatial and temporal regimes that no single platform fully addresses. Near-term milestones will be important for testing whether laboratory-scale quantum gains translate into measurable diagnostic improvements; examples include extending entanglement- and squeezing-enhanced measurement beyond the squeezed-light microscopy already demonstrated in living cells~\cite{Casacio2021,taylor2016quantum} to magnetometry of biological signals under ambient conditions, and conducting initial clinical validation of third-generation protocols. Continued progress will require coordinated advances in quantum resource engineering, device physics, and biologically informed system design, alongside careful experimental validation. The transition from energy-level readout to entanglement-enabled and learning-integrated architectures therefore reflects not only increasing quantum complexity but an expanding ambition: to move from measuring isolated observables toward extracting structured biological information under realistic clinical constraints.

\vspace{10pt} \noindent \textsf{\textbf{Acknowledgements.}} JL is supported in part by the University of Pittsburgh, School of Computing and Information, Department of Computer Science, Pitt Cyber, Pitt Momentum fund, PQI Community Collaboration Awards, John C. Mascaro Faculty Scholar in Sustainability, NSF award 2535915, Thinking Machines Lab and Cisco Research. This research used resources of the Oak Ridge Leadership Computing Facility, which is a DOE Office of Science User Facility supported under Contract DE-AC05-00OR22725.
 G.D., R.W., and J.B. are supported by National Science Foundation Award No. 2304998.
 KPS thanks the U.S. Department of Energy, Office of Science, Advanced Scientific Computing Research (ASCR) program, for support under Award Number DE-SC0026264, and PQI Community Collaboration Awards.

\let\oldaddcontentsline\addcontentsline
\renewcommand{\addcontentsline}[3]{}
\bibliographystyle{rev4-2mod}
\bibliography{bibliography}
\let\addcontentsline\oldaddcontentsline

\onecolumngrid

\setcounter{equation}{0}
\setcounter{figure}{0}
\renewcommand{\theequation}{S\arabic{equation}}
\renewcommand{\thefigure}{S\arabic{figure}}
\renewcommand{\appendixname}{}
\setlength{\tabcolsep}{3pt}
\renewcommand{\arraystretch}{1.3}
\setlength{\baselineskip}{14pt}
\linespread{1.3}

\bigskip\bigskip\bigskip

\begin{center}
{\large \bf Supplementary Material for \\ ``Four Generations of Quantum Biomedical Sensors''}
\end{center}

\tableofcontents

\appendix

\section{Quantum Sensor Timeline Table}
\label{appendix:timeline}
\begin{table}[htbp!]
\centering
\caption{Representative milestones and reported sensitivities for traditional and quantum biomedical sensors. 
Values are approximate and normalized for conceptual comparison in Fig.~1.}
\label{tab:quantum_timeline_data}
\resizebox{\textwidth}{!}{
\begin{tabular}{@{}llllll@{}}
\toprule
\textbf{Year} & \textbf{Technology / Modality} & \textbf{Representative Reference / Device} & 
\textbf{Physical Principle} & \textbf{Reported Sensitivity} & \textbf{Generation Mapping}\\
\midrule
1990s & MRI (High-field, 3\,T) & Siemens MAGNETOM & Nuclear spin resonance (ensemble) & SNR-limited (high-field); $0.47$\,fT\,Hz$^{-1/2}$ (ULF det.)~\cite{Savukov2022} & 2nd Gen \\
2000 & EEG & Standard 64-channel system & Cortical electric potential & $\sim$10$^{-6}$\,V\,Hz$^{-1/2}$ & Classical \\
2003 & SERF atomic magnetometer & Romalis group (Princeton) & Spin-exchange relaxation-free vapor & $\sim$10$^{-15}$\,T\,Hz$^{-1/2}$ (fT) & 2nd Gen \\
2008 & NV-center ODMR in diamond & Balasubramanian et al.\ (Stuttgart) & Electronic spin resonance (ODMR) & $\sim$10$^{-9}$\,T\,Hz$^{-1/2}$ (nT) & 1st Gen \\
2013 & Nanoscale NV-NMR & Staudacher et al.\ / Mamin et al. & Single-spin NMR via NV readout & $\sim$10$^{-9}$\,T\,Hz$^{-1/2}$ (nT) & 2nd Gen \\
2015 & Sub-pT NV-diamond magnetometer & Wolf et al.\ (Stuttgart) & High-density NV ensemble, optimized readout & $\sim$10$^{-12}$\,T\,Hz$^{-1/2}$ (pT) & 2nd Gen \\
2016 & Spin-squeezed atomic sensor & Hosten et al.\ (Stanford) & Entanglement-enhanced atom interferometry & ${\sim}20$\,dB below SQL & 3rd Gen \\
2020 & NV ensemble magnetometry (review) & Barry et al.\ (MIT/Harvard) & Optimized ensemble NV readout & $\sim$10$^{-12}$\,T\,Hz$^{-1/2}$ (pT) & 2nd-3rd Gen \\
2023-2025 & Quantum-AI hybrid biosensor (prototypes) & Multiple groups (conceptual) & Adaptive sensing / ML feedback & n/a (conceptual) & 4th Gen \\
\bottomrule
\end{tabular}}
\end{table}

\section{First-Generation Quantum Medical Sensor}
\subsection{Magnetoresistive biosensors (GMR/TMR)}
\label{appendix:magnetoresistive}

Magnetoresistive biosensors provide a paradigmatic example of first-generation quantum sensing in biomedical applications, where discrete spin-dependent electronic states are exploited for robust signal transduction under ambient conditions. Giant magnetoresistance sensors were first demonstrated as biosensing platforms by Baselt and colleagues, who showed that magnetically labeled DNA hybridization and protein binding events could be detected through resistance changes induced by spin-dependent scattering \cite{baselt1998biosensor}. Building on this principle, Osterfeld and colleagues developed high-density GMR nanosensor arrays capable of real-time kinetic analysis across multiple protein interactions simultaneously, enabling multiplexed diagnostics and lab-on-chip integration at clinically relevant scales \cite{Osterfeld2008}. Tunnel magnetoresistance sensors extend this paradigm by exploiting spin-dependent tunneling through thin insulating barriers, yielding substantially enhanced sensitivity. Recent implementations have integrated TMR sensors into needle-shaped magnetrodes for minimally invasive neurophysiological measurements, achieving sub-nanotesla sensitivity sufficient to resolve magnetic fields generated by individual neuronal action potentials \cite{Luo2024}. Complementary studies have demonstrated TMR-based detection of magnetic nanoparticles embedded within tissue phantoms, with direct relevance to magnetic hyperthermia monitoring where accurate spatial localization of therapeutic agents is essential \cite{ghemes2023tunnel}. Together, these results highlight the scalability and clinical relevance of magnetoresistive platforms for point-of-care diagnostics and continuous physiological monitoring.

\subsection{Quantum Dots and SET}
\label{appendix: quantum dots/SET}

Quantum dots and single-electron devices constitute a complementary class of first-generation quantum sensors operating at the nanoscale, where quantization of electronic states enables ultrasensitive molecular detection. Quantum dots have become indispensable tools for fluorescence-based biomedical imaging due to their size-tunable emission spectra and exceptional photostability. The seminal work by Chan and Nie demonstrated that bioconjugated quantum dots could serve as robust probes for cancer biomarker detection, overcoming the photobleaching limitations of conventional organic fluorophores \cite{Chan1998}. Subsequent in vivo studies by Gao and colleagues showed that peptide-functionalized near-infrared quantum dots can selectively accumulate in tumor tissue through receptor-mediated pathways, enabling high-contrast molecular imaging in living systems \cite{Gao2004}. More recent reviews have emphasized the integration of quantum dots into hybrid electrical and optical biosensing architectures, particularly for nucleic acid detection where sensitivity and miniaturization are critical \cite{Tang2025}.

Single-electron transistors offer an alternative route to quantum-enabled biosensing by exploiting charge quantization and Coulomb blockade effects. Clear Coulomb blockade requires the island charging energy to greatly exceed the thermal energy ($e^{2}/2C_{\Sigma} \gg k_{B}T$), so conventional single-island devices operate cryogenically; advances in silicon nanofabrication have nevertheless enabled room-temperature biosensors based on serially connected multi-island architectures fabricated on silicon-on-insulator substrates \cite{Nakajima2016}. These devices exhibit exceptional charge sensitivity, enabling detection of biomolecular interactions such as streptavidin binding at picomolar concentrations and immunodetection of prostate-specific antigen at clinically relevant thresholds \cite{Nakajima2013}. A macroscopic implementation using porous silicon Schottky junctions further demonstrated that single-electron effects can be preserved across centimeter-scale areas, opening pathways toward low-cost, large-area biosensor arrays suitable for clinical deployment \cite{Ashoori2021}. When integrated with microfluidic and organ-on-chip platforms, these nanoscale quantum sensors offer promising routes toward real-time monitoring of cellular responses and personalized therapeutic screening.

\section{Supplementary Sensor Platform Specifications and Readiness Assessment}
\label{appendix: specification}

\begin{table*}[htbp]
\caption{Dual-axis cross-platform comparison: physics performance versus translational readiness
for representative quantum and conventional biomedical sensing platforms.
$T_{\mathrm{op}}$: sensor operating temperature;
SWaP-C: size, weight, power, and cost.
Gen.\ column: quantum-technology generation (1st-4th); ``n/a'' denotes a classical/conventional platform.}
\label{tab:dualaxis}
\centering
\scriptsize
\begin{tblr}{
  width = \textwidth,
  colspec = {X[1.4,l] X[0.4,c] X[1.3,l] X[1.2,l] X[0.9,l] X[0.7,c] X[1.1,l] X[1.1,l] X[1.5,l] X[1.2,l] X[1.0,l]},
  cells = {font=\scriptsize\RaggedRight},
  row{1} = {bg=headerblue, fg=white, font=\bfseries\scriptsize, valign=m, halign=c},
  row{2,4,6,8,10,12,14} = {bg=rowalt},
  rowsep = 2pt, colsep = 2pt,
  hline{1,Z} = {0.08em},
  hline{2} = {0.05em},
}
Platform & Gen. & Sensitivity & Bandwidth & Spatial Res. & {$T_{\mathrm{op}}$\\(K)} & {Sensor-\\Sample Dist.} & {Shielding\\Req.} & SWaP-C & Biocompat. & Reg.\ Path \\
MRI/fMRI & 2nd & {$0.47$\,fT/\\$\sqrt{\mathrm{Hz}}$\textsuperscript{$\dagger$}} & MHz (RF) & 0.5-2\,mm & {${\sim}4$\\(SC)} & 5\,mm-2\,cm & {Faraday\\cage} & {${>}$\$1\,M, tons,\\${>}50$\,kW} & {External\\(Gd toxic.)} & Cleared (II) \\
PET/CT & n/a & {Picomolar\\(mol.)} & {n/a\\(static)} & 4-5\,mm & RT & {External\\ring} & {None\\(mag.)} & {${>}$\$1\,M,\\cyclotron} & {Ionizing\\rad.} & Cleared (II) \\
EEG & n/a & {${\sim}\upmu$V\\(potential)} & {DC-\\hundreds\,Hz} & 10-25\,mm & RT & {Scalp\\contact} & None & {${<}$\$50\,k,\\portable} & {Skin\\contact} & Cleared (II) \\
SQUID-MEG & 2nd & {2-5\,fT/\\$\sqrt{\mathrm{Hz}}$} & {DC-\\${\sim}1$\,kHz} & 5-10\,mm & {4.2\\(LHe)} & {2-3\,cm\\(dewar)} & {Multi-layer\\MSR} & {\$2-4\,M,\\${>}400$\,kg} & External & 510(k) \\
{OPM-MEG\\(SERF)} & 2nd & {7-15\,fT/\\$\sqrt{\mathrm{Hz}}$} & {DC-\\${\sim}150$\,Hz} & 3-5\,mm & {393-\\473} & 3-6\,mm & {MSR +\\nulling} & {\$50-200\,k,\\wearable} & {Skin\\(insul.)} & {510(k)\\pred.} \\
{NV diamond\\(ens.)} & 1-2 & {1-10\,nT/$\sqrt{\mathrm{Hz}}$;\\${\sim}$pT pulsed} & kHz-GHz & {${\sim}10\,\upmu$m (WF);\\nm} & {288-\\333} & {${<}1$\,mm-\\$10\,\upmu$m} & None & {${<}$\$10\,k;\\miniature} & {Diamond\\(inert)} & De Novo \\
{FND (nano-\\diamond)} & 1st & Single-spin & {MHz\\(relax.)} & {nm\\(intracell.)} & RT & Zero & None & Minimal & Debated & De Novo \\
{Quantum\\dots} & 1st & {Molecular\\(fluor.)} & {n/a\\(steady)} & ${\sim}10$\,nm & RT & {Zero\\(label)} & None & Low cost & {Cd: toxic;\\C: good} & {Material-\\dep.} \\
{Hyperpol.\\$^{13}$C MRI} & 2nd & {${>}10^{4}\times$\\enh.} & {MHz\\(MRI)} & ${\sim}$mm & {${\sim}4$\\+ RT} & {cm\\(deep)} & {MRI\\built-in} & {Large: MRI\\+ pol.} & {No\\radiation} & {510(k)\\ext.} \\
{Rydberg\\atom} & 2nd & {$\upmu$V-nV/cm/\\$\sqrt{\mathrm{Hz}}$} & GHz-THz & ${\sim}$mm & {298-\\353} & {mm-cm\\{\textsuperscript{$\ast$}}} & {Vacuum\\cell} & {Medium,\\optical} & Incompat. & Unclear \\
{Entangled\\NV pairs} & 3rd & {${\sim}5$\,dB\\${>}$ SQL} & kHz & ${<}100$\,nm & RT & ${<}100$\,nm & None & Lab-scale & {Same as\\NV} & Long-term \\
{Spin-squeezed\\vapor} & 3rd & {${\sim}2$\,dB\\${>}$ SQL} & DC-kHz & mm-cm & {350-\\470} & mm-cm & MSR & {OPM\\+ ent.} & {Same as\\OPM} & Long-term \\
{Distributed\\QSN} & 4th & {Heisenberg\\$1/N$} & Adaptive & {Multi-\\node} & Heterog. & {Cross-\\organ} & TBD & {QP +\\transd.} & TBD & ${>}10$\,yr \\
\end{tblr}

\vspace{4pt}
{\scriptsize
\noindent\textsuperscript{$\dagger$}Ultra-low-field MRI detection-coil sensitivity (2.8\,mT, room-temperature high-sensitivity coil)~\cite{Savukov2022}; conventional high-field receive coils are specified by SNR rather than field-equivalent noise.\\
\textsuperscript{$\ast$}The atoms are enclosed in a vapor cell, so the minimum sensor-sample standoff is set by the cell-wall thickness: microfabricated thin-walled cells reach ${\sim}$0.1-1\,mm~\cite{Kbler2010,Zhang2018}, whereas the centimeter-scale glass cells used in current Rydberg electrometry demonstrations place the atoms ${\sim}$1\,cm from an external sample~\cite{sedlacek2012microwave,holloway2014broadband}.
}

\end{table*}

\begin{table*}[htbp]
\caption{Key bottlenecks of existing medical detection devices and corresponding quantum sensing breakthrough directions.
For each established clinical modality, the table identifies core limitations, their physical origins,
and specific quantum sensor platforms that address each bottleneck.}
\label{tab:bottleneck}
\centering
\scriptsize
\begin{tblr}{
  colspec = {p{1.4cm} p{2.6cm} p{3.8cm} p{3.0cm} p{4.0cm}},
  row{1} = {bg=headerblue, fg=white, font=\bfseries\scriptsize, valign=m},
  row{2,4,6} = {bg=rowalt},
  rowsep = 1.5pt, colsep = 2.5pt,
  hline{1,Z} = {0.08em},
  hline{2} = {0.05em},
}
Existing Device & Core Strengths & Key Bottlenecks / Unmet Needs & Physical Origin of Bottleneck & Quantum Sensor Breakthrough Direction \\
MRI (1.5-7\,T)
& Deep-tissue imaging (cm-scale imaging depth); soft-tissue contrast; no ionizing radiation; multi-sequence
& (i) Scan 20-60\,min, immobility; (ii) fMRI BOLD delay ${\sim}5$\,s; (iii) thermal polarization low (${\sim}10^{-5}$); (iv) Gd nephrotoxicity; (v) ${>}$\$1M, LHe-dep.; (vi) implant contraindications
& Thermal spin polarization very low; BOLD is indirect proxy; SC magnet needs 4\,K; SAR limits
& Hyperpol.\ $^{13}$C: ${>}10^{4}\times$ signal $\to$ metabolic imaging; OPM $\to$ LHe-free MRI; NV nanoscale NMR; quantum phase estimation $\to$ accelerated acquisition \\
PET (FDG)
& Picomolar sensitivity; whole-body metabolic; tumor staging gold standard
& (i) Ionizing radiation (pediatric); (ii) 4-5\,mm resolution; (iii) cyclotron needed; (iv) short tracer half-life; (v) low early-stage sensitivity; (vi) inflammation false positives
& $e^{+}$ annihilation $\to$ 511\,keV $\gamma$ $\to$ inherent radiation; detector geometry; half-life fixed
& QD $\to$ radiation-free labeling; NV relaxometry $\to$ immunomagnetic; hyperpol.\ MRI $\to$ replace some PET; quantum-enhanced photoacoustic \\
CT
& Fast (${<}1$\,min); high spatial resolution; bone/lung; emergency gold standard
& (i) Ionizing radiation (cancer risk); (ii) poor soft-tissue contrast; (iii) no metabolic info; (iv) iodine contrast toxicity
& X-ray absorption $\to$ good for bone/air, poor soft tissue; ionizing radiation inherent
& Quantum illumination $\to$ low-dose high-contrast (theoretical); entanglement-enhanced tomography $\to$ reduced photon requirement \\
EEG
& ms temporal; low cost (${<}$\$50k); portable; non-invasive
& (i) Poor spatial resolution (cm); (ii) skull distortion (error 25\,mm); (iii) EMG/EOG artifacts; (iv) low deep-source sensitivity; (v) no psychiatric biomarker
& Skull $\to$ potential distortion; superposition $\to$ ill-posed inverse problem
& MEG (SQUID/OPM) $\to$ $B$ unaffected by skull; wearable OPM-MEG; OPM+EEG $\to$ super-additive source model \\
SQUID-MEG
& fT sensitivity; ms temporal; mm localization; adjunctive presurgical epilepsy mapping
& (i) LHe cryogenics $\to$ extreme cost; (ii) rigid helmet; (iii) ${\sim}200$ units globally; (iv) 2-3\,cm standoff; (v) He volatility ${>}400$\%
& Josephson junctions need ${<}4.2$\,K $\to$ LHe dewar $\to$ fixed standoff; He non-renewable
& OPM: LHe-free $\to$ wearable $\to$ 3-5$\times$ gain; high-$T_{\mathrm{c}}$ SQUID; NV arrays (long-term) \\
Tumor biomarkers (PSA, CEA, etc.)
& Blood test: simple, low cost, mass screening
& (i) Low early-stage sensitivity; (ii) poor specificity; (iii) tumor heterogeneity; (iv) healthy-cell background; (v) ctDNA: 7\% Stage~I lung
& Small tumors release minimal markers; concentration in physiological noise; single-analyte is low-dimensional
& SET/QD $\to$ single-molecule LOD; NV relaxometry $\to$ wash-free immunoassay; quantum ML $\to$ pattern recognition; multiplexed quantum arrays \\
\end{tblr}
\end{table*}

\end{document}